\documentclass[fleqn,usenatbib]{mnras}

\usepackage{newtxtext,newtxmath}

\usepackage[T1]{fontenc}

\DeclareRobustCommand{\VAN}[3]{#2}
\let\VANthebibliography\thebibliography
\def\thebibliography{\DeclareRobustCommand{\VAN}[3]{##3}\VANthebibliography}

\usepackage{graphicx}	
\usepackage{amsmath}	

\newcommand{\fermi}{{\it Fermi}-LAT}
\newcommand{\gray}{$\gamma$-ray}
\newcommand{\grays}{$\gamma$-rays}
\newcommand{\source}{3C 454.3}

\title[Multiwavelength emission from 3C454.3]{Modeling the Broadband Emission of 3C 454.3}

\author[N. Sahakyan]{
N. Sahakyan,$^{1,2, 3}$ \thanks{E-mail: narek@icra.it}
\\
$^{1}$ICRANet-Armenia, Marshall Baghramian Avenue 24a, Yerevan 0019, Armenia\\
$^{2}$ICRANet, P.zza della Repubblica 10, 65122 Pescara, Italy\\
$^{3}$ ICRA, Dipartimento di Fisica, Sapienza Universita` di Roma, P.le Aldo Moro 5, 00185 Rome, Italy\\
}

\date{Accepted XXX. Received YYY; in original form ZZZ}

\pubyear{2015}

\begin{document}
\label{firstpage}
\pagerange{\pageref{firstpage}--\pageref{lastpage}}
\maketitle

\begin{abstract}
The results of a long-term multiwavelength study of the powerful flat spectrum radio quasar 3C 454.3 using {\it Fermi}-LAT and Swift XRT/UVOT data are reported. In the $\gamma$-ray band, {\it Fermi}-LAT observations show several major flares when the source flux was $>10^{-5}\:{\rm photon\:cm^{-2}\:s^{-1}}$; the peak $\gamma$-ray flux above $141.6$ MeV, $(9.22\pm1.96)\times10^{-5}\:{\rm photon\:cm^{-2}\:s^{-1}}$ observed on MJD 55519.33, corresponds to $2.15\times10^{50}\:{\rm erg\:s^{-1}}$ isotropic $\gamma$-ray luminosity. The analysis of Swift XRT and UVOT data revealed a flux increase, although with smaller amplitudes, also in the X-ray and optical/UV bands. The X-ray emission of 3C 454.3 is with a hard spectral index of $\Gamma_{\rm X}=1.16-1.75$, and the flux in the flaring states increased up to $(1.80\pm0.18)\times10^{-10}{\rm erg\:cm^{-2}\: s^{-1}}$. Through combining the analyzed data, it was possible to assemble 362 high-quality and quasi-simultaneous spectral energy distributions of 3C 454.3 in 2008-2018 which all were modeled within a one-zone leptonic scenario assuming the emission region is within the broad line region, involving synchrotron, synchrotron self-Compton and external Compton mechanisms. Such an extensive modeling is the key for constraining the underlying emission mechanisms in the 3C 454.3 jet and allows to derive the physical parameters of the jet and investigate their evolution in time. The modeling suggests that during the flares, along with the variation of emitting electron parameters, the Doppler boosting factor increased substantially implying that the emission in these periods has most likely originated in a faster moving region.
\end{abstract}




\section{Introduction}
\indent Blazars are a subclass of active galactic nuclei with a jet pointing or making a small angle with respect to the observer \citep{1995PASP..107..803U}. These jets are strong sources of electromagnetic radiation ranging from radio to high energy (HE; $>100$ MeV) and very high energy \gray\ bands (see \citealt{2017A&ARv..25....2P} for a recent review). This emission is characterized by rapid and high-amplitude variability in almost all wavelengths, the most extreme being at \gray\ band \citep[e.g., order of minutes, ][]{2016ApJ,foschini11,foschini13, nalewajko, brown13, rani13,saito,hayashida15, 2018ApJ...854L..26S}. Blazars are dominant sources in the extragalactic \gray\ sky and have been observed even at very high redshifts \citep[e.g.,][]{2017ApJ...837L...5A, 2020MNRAS.498.2594S}. The recent studies of 4FGL J1544.3-0649 reveal the possible existence of transient blazars; being undecteable in the X-ray and \gray\ bands, 4FGL J1544.3-0649 for a few months rose to be one of the brightest known X-ray blazars \citep{2021MNRAS.502..836S}. If 4FGL J1544.3-0649 does not indeed represent an isolated case but rather a common phenomenon, this would have a non-negligible role in the multimessenger astrophysics.\\
\indent The broadband spectral energy distribution (SED) of blazars is characterized by two broad humps, one at optical/UV/X-ray bands and the other in the HE \gray\ band 
(see \citealt{2017A&ARv..25....2P} for a recent review). It is believed that the first peak (low energy component) is mostly due to synchrotron emission from relativistic electrons, whereas the origin of the second component is highly debatable. Within conventional leptonic scenarios, this component is produced when the synchrotron emitting electrons inverse Compton up scatter the photons of internal \citep[synchrotron self Compton (SSC)][]{ghisellini, bloom, maraschi} 
or external \citep[external inverse Compton (EIC)][]{blazejowski,ghiselini09, sikora} origin. The nature of the external photon fields depends on the distance of the emitting region from the central black hole \citep{2009ApJ...704...38S} and can be dominated either by the photons directly emitted from the accretion disk \citep{1993ApJ...416..458D, 1992A&A...256L..27D} or disk photons reflected from the broad-line region \citep[BLR;][]{sikora} or IR photons emitted from the dusty torus \citep{blazejowski}. Recently, after associating TXS 0506+056 with the IceCube-170922A neutrino event \citep{2018Sci...361..147I, 2018Sci...361.1378I, 2018MNRAS.480..192P}, it is more evident that the HE component could be initiated by the interaction of energetic protons when they are effectively accelerated in the blazar jets. The HE component can be either from proton synchrotron emission \citep{2001APh....15..121M} or from secondary particles from pion decay \citep{1993A&A...269...67M, 1989A&A...221..211M, 2001APh....15..121M, mucke2, 2013ApJ...768...54B}. In the latter case, blazars are also sources of very high energy neutrinos \citep{2018ApJ...863L..10A,2018ApJ...864...84K, 2018ApJ...865..124M, 2018MNRAS.480..192P, 2018ApJ...866..109S, 2019MNRAS.484.2067R,2019MNRAS.483L..12C, 2019A&A...622A.144S, 2019NatAs...3...88G}.\\
\indent Based on the properties observed in the optical band, blazars are classified as Flat Spectrum Radio Quasars (FSRQs) when the emission lines are stronger and quasar-like or BL Lacs when these lines are weak or absent \citep{1995PASP..107..803U}. Alternatively, blazars are FSRQs when the luminosity of the broad emission lines (or accretion disk) measured in Eddington units is $L_{\rm BLR}/L_{\rm Edd}\geq 5\times10^{-4}$, otherwise they are BL Lacs \citep{2011MNRAS, 2012MNRAS}. Depending on the position of the synchrotron component peak ($\nu_{\rm s}$), the blazars are further classified as low synchrotron peaked (LSP) sources, when $\nu_{\rm s}<10^{14}$ Hz, intermediate synchrotron peaked (ISP) and high synchrotron peaked (HSP) sources when $10^{14}<\nu_{\rm s}<10^{15}$ Hz and $\nu_{\rm s}>10^{15}$ Hz, respectively \citep{Padovani1995,Abdo_2010}. However, sometimes the synchrotron peak of HSPs can reach $\sim$1 keV, ($\sim 2\times 10^{17}$ Hz) or beyond, showing an extreme behaviour \citep[e.g.][]{sedentary,2001A&A...371..512C,Biteau2000}. Such behaviour was first observed during the flare of Mkn 501 when the synchrotron peak reached $\sim 100$ keV \citep{1998ApJ...492L..17P}, and then many such objects were identified in the X-ray observations. For example, during the flares of 1ES 1218+304 the X-ray spectral index hardened to $\Gamma \leq1.80$, shifting the peak towards higher energies \citep{2020MNRAS.496.5518S}. In this classification, FSRQs usually have a synchrotron peak at $\nu_{\rm s}<10^{14}$ Hz, so they are LSPs.\\
\indent The BL Lac and FSRQ SEDs demonstrate different properties. In FSRQs, the strong external photon fields 
which are weak or absent in the case of BL Lacs, modify the HE component: these external photons are seen relativistically boosted in the comoving frame of the jet and can dominate over the internal synchrotron photon fields, giving rise to the EIC component \citep[e.g.,][]{2018ApJ...863..114G, 2017MNRAS.470.2861S}. Therefore, in FSRQs the luminosity of the second component is usually larger, i.e. shows a larger Compton dominance \citep{sikora} due to the presence of external seed photons. The shape of this component depends on the distribution of up-scattering photons which in its turn is defined by the distance of the emitting region from the central black hole. So, the modeling of the SED or features in the \gray\ spectrum (e.g., break or cut-off) can help to localize the emission region
\citep[e.g.,][]{2010ApJ...717L.118P, 2011ApJ...730L...8A, 2020A&A...635A..25S}.\\
\indent \source\ is a typical FSRQ at $z=0.859$ harboring a black hole with a mass estimated to be $1.5\times10^9\:M_\odot$ \citep{2002ApJ...579..530W, 2006ApJ...637..669L}. This source was extensively studied in the multiwavelength band over the last two decades \citep[e.g.,][]{2006A&A...456..911G}. However, after the prominent optical outburst in 2005 with a peak optical brightness of $R=12$ mag, the source remained active showing several bright flares, it has become a target of multiwavelength studies \citep{2008A&A...491..755R, 2011A&A...534A..87R, 2010ApJ...716L.170P, 2011ApJ...733L..26A, 2009ApJ...699..817A, 2010ApJ...721.1383A, 2013ApJ...773..147J, 2021ApJ...906....5A}. This multiwavelength campaigns provided unprecedented information on this source. For example, \citet{2009ApJ...697L..81B} studying the multiwavelength data observed during the high flux state in July 2008, showed that the emissions in IR, optical, UV, and \gray\ bands are well correlated while the X-ray flux is correlated with none of these fluxes. These features in the multiwavelength variability can be naturally explained within the EIC scenario. However, \citet{2011MNRAS.410..368B} found that the optical, X-ray and \gray\ fluxes correlate during the extreme brightening of \source\ in the first week of December 2009. In the \gray\ band the source was first detected by EGRET on the Compton Gamma-Ray Observatory \citep{1993ApJ...407L..41H} and then extensively monitored by AGILE and  Fermi large area telescope (\fermi) since 2007. The AGILE observations have shown that the \source\ was in active \gray\ state in several occasions \citep{2009ApJ...707.1115D, 2010ApJ...716L.170P, 2011ApJ...736L..38V}. In 2010 during the exceptional bright \gray\ flare \fermi\ detected a flux above $\sim10^{-5}\:{\rm photon\:cm^{-2}\:s^{-1}}$ from \source\ which corresponds to $\sim10^{50}\:{\rm erg\:s^{-1}}$ isotropic luminosity, making \source\ one of the brightest \gray\ sources in the sky \citep{2011ApJ...733L..26A}. During the flaring of \source, different extraordinary changes in terms of spectral and temporal properties were reported in the multiwavelength context and especially in the \gray\ band. For example, \citet{2009ApJ...699..817A} found that the \gray\ spectrum of \source\ steepens above $\sim2$ GeV and it is better explained by a broken power-law with photon indices of $\sim2.3$ and $\sim3.5$ below and above the break, respectively. This turnover in the spectrum was interpreted by the spectral break in the electron distribution; however, \citet{2010ApJ...714L.303F} interpreted the break to be due to the inverse Compton scattering of accretion disk and BLR photons.\\
\indent In this paper, the evolution of the multiwavelength SED of \source\ during 2008-2018 as well as the spectral changes in different bands are investigated. These changes in the SED are expected to arise from the variation of the parameters of the emitting electrons or the physical parameters of the emission region \citep{2011ApJ...736..128P}, so, they directly define the physical processes taking place in the jet as well as the state of the plasma in it. For this reason, in the period from 2008 to 2018, as many SEDs of \source\ as possible that can be constructed with contemporaneous data from the radio to HE bands have been modeled within a leptonic scenario. More explicitly, these SEDs are modeled assuming the low energy component is due to synchrotron emission whereas the second component is due to SSC and/or EIC by the same electrons. Through such modeling the jet parameters are estimated for different periods allowing to investigate their variation which
could help to investigate the changes in the jet of \source\ as well as to understand the origin of the flares. \source\ was selected because of \textit{i)} the availability of rich multiwavelength data (e.g., more than 465 observations with Neil Gehrels Swift Observatory (hereafter Swift) and continuous monitoring by \fermi) and  \textit{ii)} its large amplitude variability in almost all wavelengths.\\
\indent In the current paper the origin of the multiwavelength emission from \source\ is investigated using the multiwavelength data accumulated during 2008-2018. The paper is structured as follows. Section \ref{datanal} presents the multiwavelength data analyzed in the current study. The evolution of SED in time is presented in Section \ref{sedevol}. The modeling of SEDs is presented in Section \ref{ome} and the discussions in Section \ref{DC}. The conclusion is given in Section \ref{conc}.
\begin{figure*}
	\includegraphics[width=0.9\textwidth]{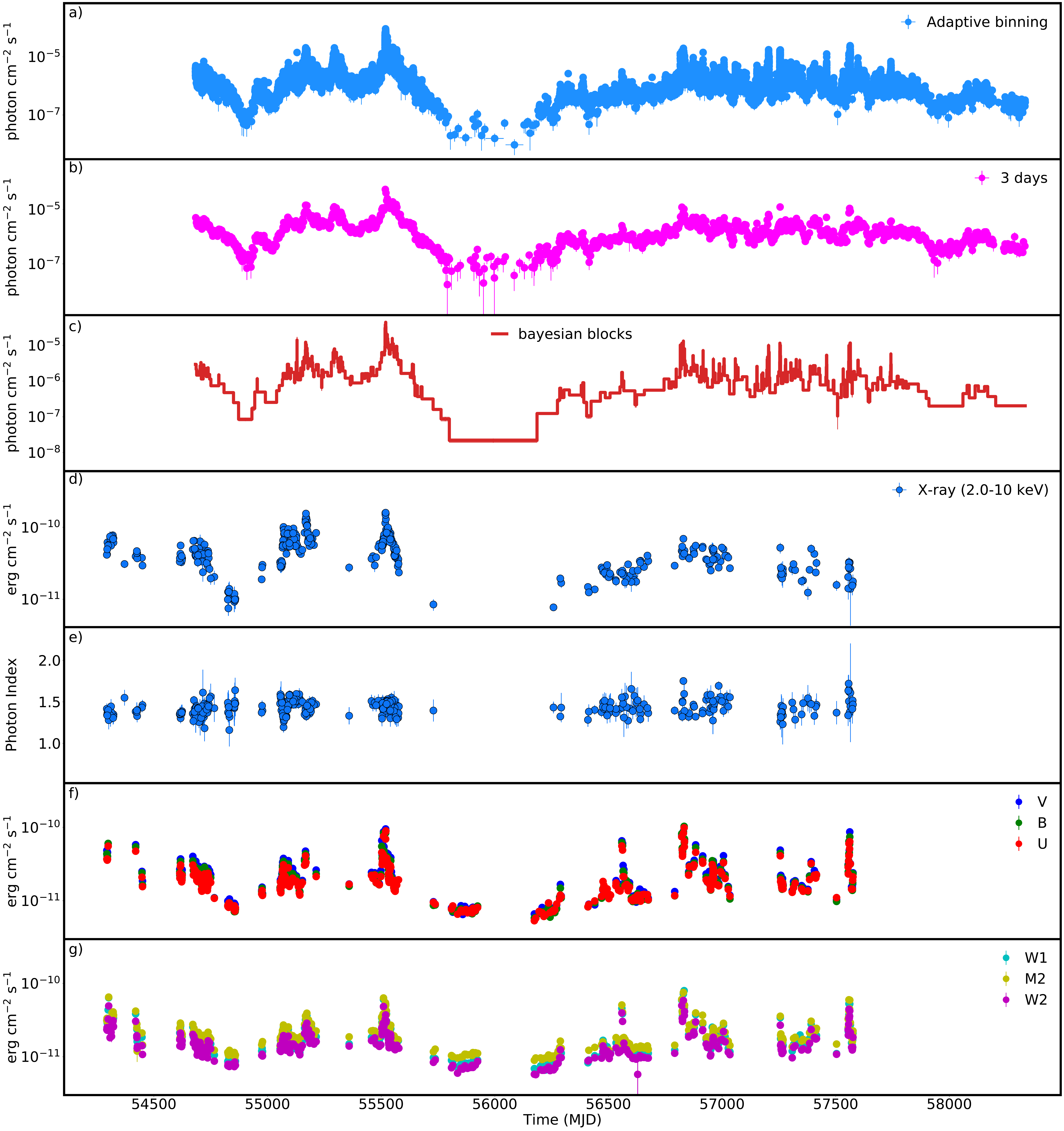}
    \caption{The multiwavelength light curve of \source\ between May 01, 2007 and August 04, 2018. From top to bottom: adaptively binned \gray\ light curve ($>141.6$ MeV), 3-day binned \gray\ light curve ($>100$ MeV), Bayesian block representation of the \gray\ light curve, 2.0-10 keV X-ray flux, 0.3-10.0 keV X-ray photon index, flux in V, B, and U filters and flux in  W1, M2 and W2 filters.}
    \label{lightcurve_all}
\end{figure*}
\section{Multiwavelenth Observations}\label{datanal}
\source, being among the brightest \gray\ blazars, was frequently monitored in different energy bands. Here, the emission of \source\ in the optical/UV, X-ray and \gray\ bands is investigated using the data from \fermi, Swift UVOT and XRT telescopes.
\subsection{\fermi\ data}
\fermi\ on board the Fermi Gamma-ray Space Telescope is a pair-conversion telescope sensitive to \grays\ in the energy band from $20$ MeV to  $500$ GeV. Scanning the entire sky every $\sim3$ hours, it provides the deepest view of the \gray\ sky \citep{2009ApJ...697.1071A}.\\
\indent In the current study, the publicly available \fermi\ data accumulated between August 4, 2008 and August 4, 2018 are used. The Pass 8 (P8R3) \fermi\ events in the energy range from 100 MeV to 500 GeV extracted from a $12^\circ$ region of interest (ROI) around the \gray\ position of \source\ (RA $= 343.497$ and Dec $= 16.149$) have been analyzed using Fermi ScienceTools (1.2.1) and the P8R3\_ SOURCE\_ V2 instrument response functions (IRFs). With the help of {\it gtselect} tool the front and back events of type 3 and event class $128$ coming from zenith angles smaller than $90^{\circ}$, to reduce contamination by photons from Earth's atmosphere, were selected. Instead, the good time intervals are selected with {\it gtmktime} tool using the filter expression (DATAQUAL $>$0) and (LAT CONFIG$==$1). The events are binned within a $16.9^{\circ}\times16.9^{\circ}$ square region with a stereographic projection into $0.1^{\circ} \times 0.1^{\circ}$ pixels and into 37 equal logarithmically spaced energy bins with the help of {\it gtbin} tool. Then, an exposure map in the ROI with $22^{\circ}$ radius was computed using tasks {\it gtltcube} and {\it gtexpmap}. The background point sources from the \fermi\ fourth source catalog \citep[4FGL;][]{2020ApJ...892..105A} within ROI+5 from the position of \source\ were all included in the model file with the same spectral models as in the catalog. The normalization and spectral parameters of the sources within the ROI were set as free parameters, while that of the sources outside ROI were fixed to the catalog values. The model file contains the standard templates describing the diffuse emission from the Galaxy (gll\_ iem\_ v07) and the isotropic \gray\ background (iso\_ P8R3\_ SOURCE\_ V2\_ v1). The normalization of both components is considered as a free parameter in the analysis. Initially, the binned likelihood analysis is applied to the full time data set using the {\it gtlike} tool.\\
\indent In the light curve calculations (shorter periods), the flux and photon index are estimated applying unbinned likelihood analysis. The photon indexes of all sources except \source\ are fixed, only keeping free the normalization of the sources within the ROI. As no variability is expected from background models, their normalization was fixed. Since the likelihood fitting is performed for short periods, the spectrum of \source\ was modeled as a power-law (PL) with the normalization and photon index as free parameters. The significance of the source emission in each interval is evaluated using test statistics defined as $TS = 2log(L_1/L_0)$, where $L_0$ and $L_1$ are the likelihoods of the model without source (null hypothesis) and the alternative likelihood (with source), respectively.\\
\indent \source\ is a well known strongly variable \gray\ blazar, so the light curve is computed in two different ways. Initially, the light-curve with 3-day binning was calculated (Fig. \ref{lightcurve_all} panel b), but at this fixed time binning the fast variation of the flux will be smoothed out and the true increase or the variation of the flux cannot be investigated. In the case of short time intervals, the flux can be estimated only in the active state when the source is bright. Thus, in order to have a deeper and detailed view of the \gray\ flux variation, the light curve was generated with the help of the adaptive binning method \citep{2012A&A...544A...6L}. In this case, the bin width is defined by requiring a constant relative flux uncertainty, so the time bins are longer during low flux levels and narrower when the source is in flaring state. The light curve generated by this method is a powerful tool for investigation of the flux variations in short time scales and identification of flaring periods as well as it contains maximum possible information on the flux variation \citep[e.g., see][]{2018ApJ...863..114G, 2018A&A...614A...6S, 2017A&A...608A..37Z, 2017ApJ...848..111B, 2016ApJ...830..162B, 2013A&A...557A..71R}.\\
\indent The adaptively binned light curve with 20\% uncertainty and above $E_0 = 141.6$ MeV  \citep[for $E_0$ see][]{2012A&A...544A...6L} is in Fig. \ref{lightcurve_all} panel \textit{a)} showing the complex behavior of \source\ in 2008-2018. The source is so bright that the flux and photon index were estimated in 11698 time intervals providing a possibility to investigate the \gray\ flux changes also in hour scales. 
\begin{figure}
	\includegraphics[width=0.49\textwidth]{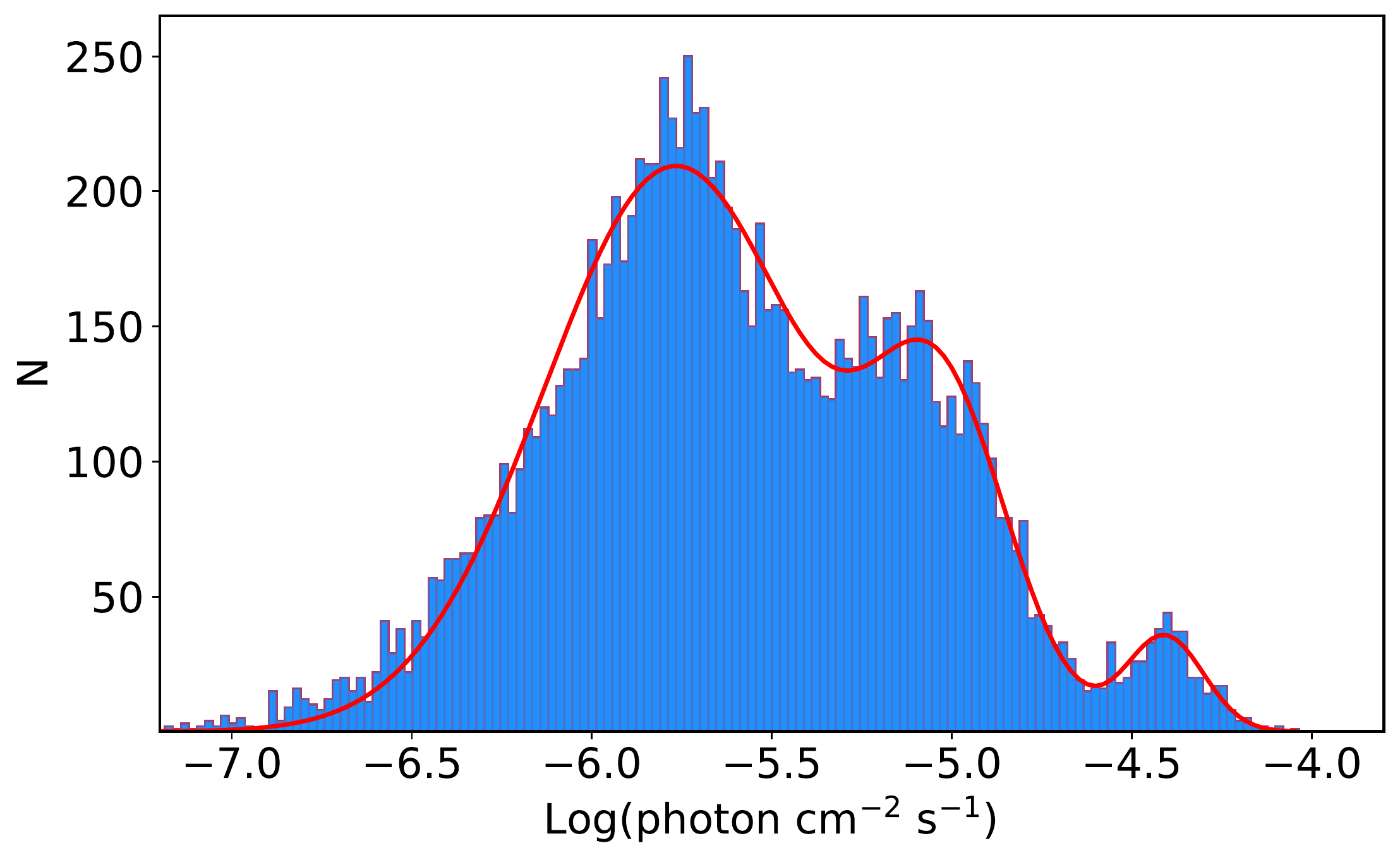}
    \caption{The distribution of the \gray\ flux estimated in adaptively binned intervals. The red line shows the fit with three Gaussian functions.}
    \label{histo}
\end{figure}
The source exhibited several substantial \gray\ flaring events during the considered period. In the low state (e.g., MJD 55800-56140) the \gray\ flux is of the order of $\simeq5\times10^{-8}\:{\rm photon\:cm^{-2}\:s^{-1}}$, however, during the flares the flux was above $10^{-5}\:{\rm photon\:cm^{-2}\:s^{-1}}$. 
The major \gray\ flaring activity was observed during MJD 55517-55522 when the highest \gray\ flux of $(9.22\pm1.96)\times10^{-5}\:{\rm photon\:cm^{-2}\:s^{-1}}$ above 141.6 MeV was observed on MJD 55519.3 with $15.6\sigma$ significance which is $1844$ times higher than the lowest \gray\ flux. Interestingly, there are in total 1657 time intervals when the source flux was above $10^{-5}\:{\rm photon\:cm^{-2}\:s^{-1}}$ which were observed in different periods, showing that from time to time the source was in a powerful \gray\ emitting state. Fig. \ref{histo} shows the distribution of the \gray\ flux of \source\ estimated in 11698 intervals. The distribution shows three peaks which characterize different states of the source. The fit of these peaks with three Gaussian functions is shown with red line in Fig. \ref{histo}. The first peak is at $1.71\times10^{-6}\:{\rm photon\:cm^{-2}\:s^{-1}}$ which is the average \gray\ flux of the source. The other two peaks are at $9.32\times10^{-6}\:{\rm photon\:cm^{-2}\:s^{-1}}$ and $3.9\times10^{-5}\:{\rm photon\:cm^{-2}\:s^{-1}}$, respectively, when the source was in an active \gray\ emitting state. Especially interesting are the periods when the flux exceeded the third peak; for example, in 72 intervals the source was in a hyperactive state when the \gray\ flux varied within $(5.01-9.22)\times10^{-5}\:{\rm photon\:cm^{-2}\:s^{-1}}$, corresponding to an apparent isotropic \gray\ luminosity of $(0.88-5.69)\times10^{50}\:{\rm erg\:s^{-1}}$, which is the range of the highest \gray\ luminosity observed from blazars so far. It is interesting to note that the \source\ flux was $>10^{-5}\:{\rm photon\:cm^{-2}\:s^{-1}}$ for 30.8 days in total ($\sim0.84$ \% of the considered ten years). This once more confirms that \source\ is extremely variable in the \gray\ band.

\begin{figure*}
	\includegraphics[width=0.47\textwidth]{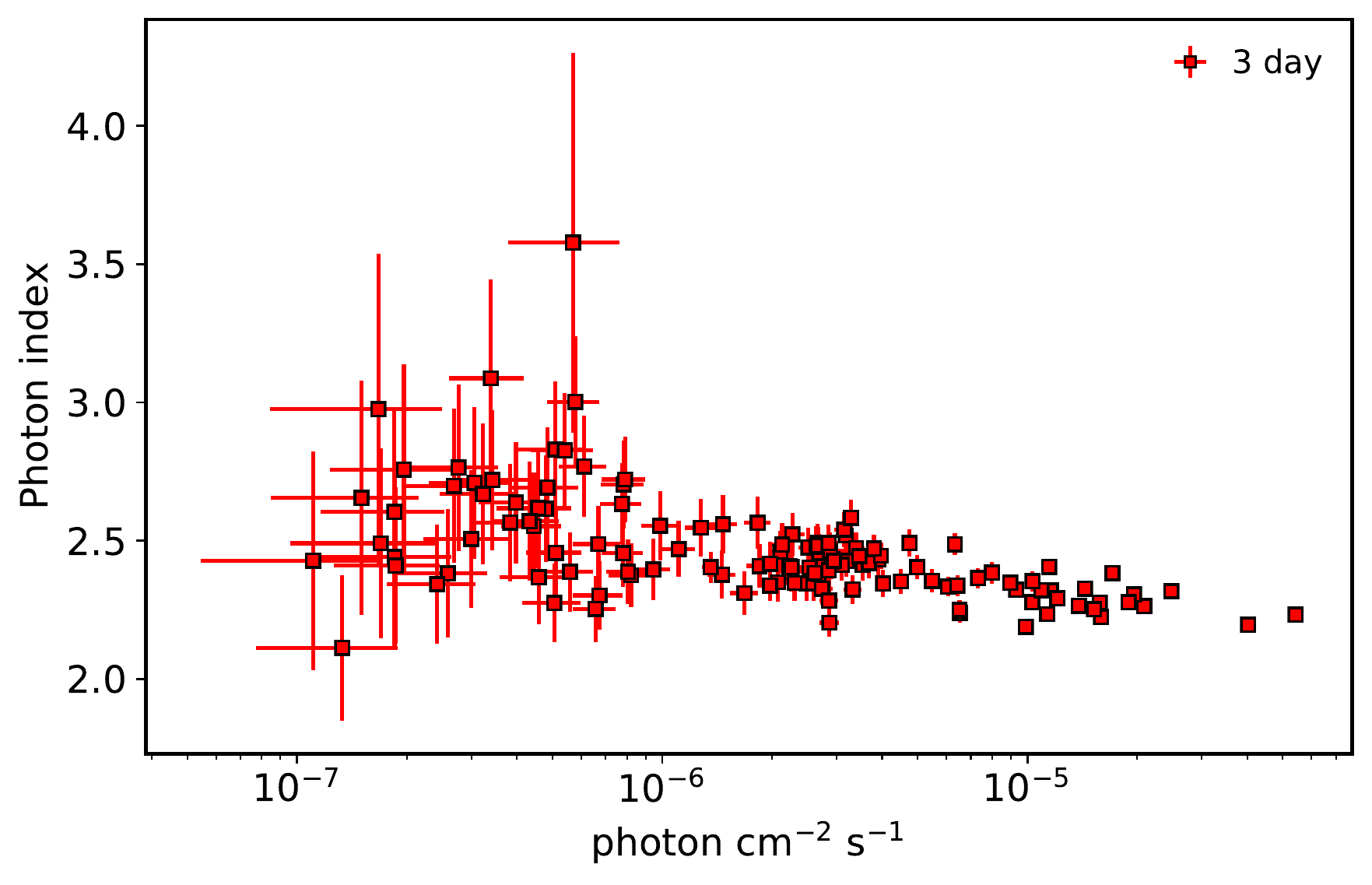}
	\includegraphics[width=0.47\textwidth]{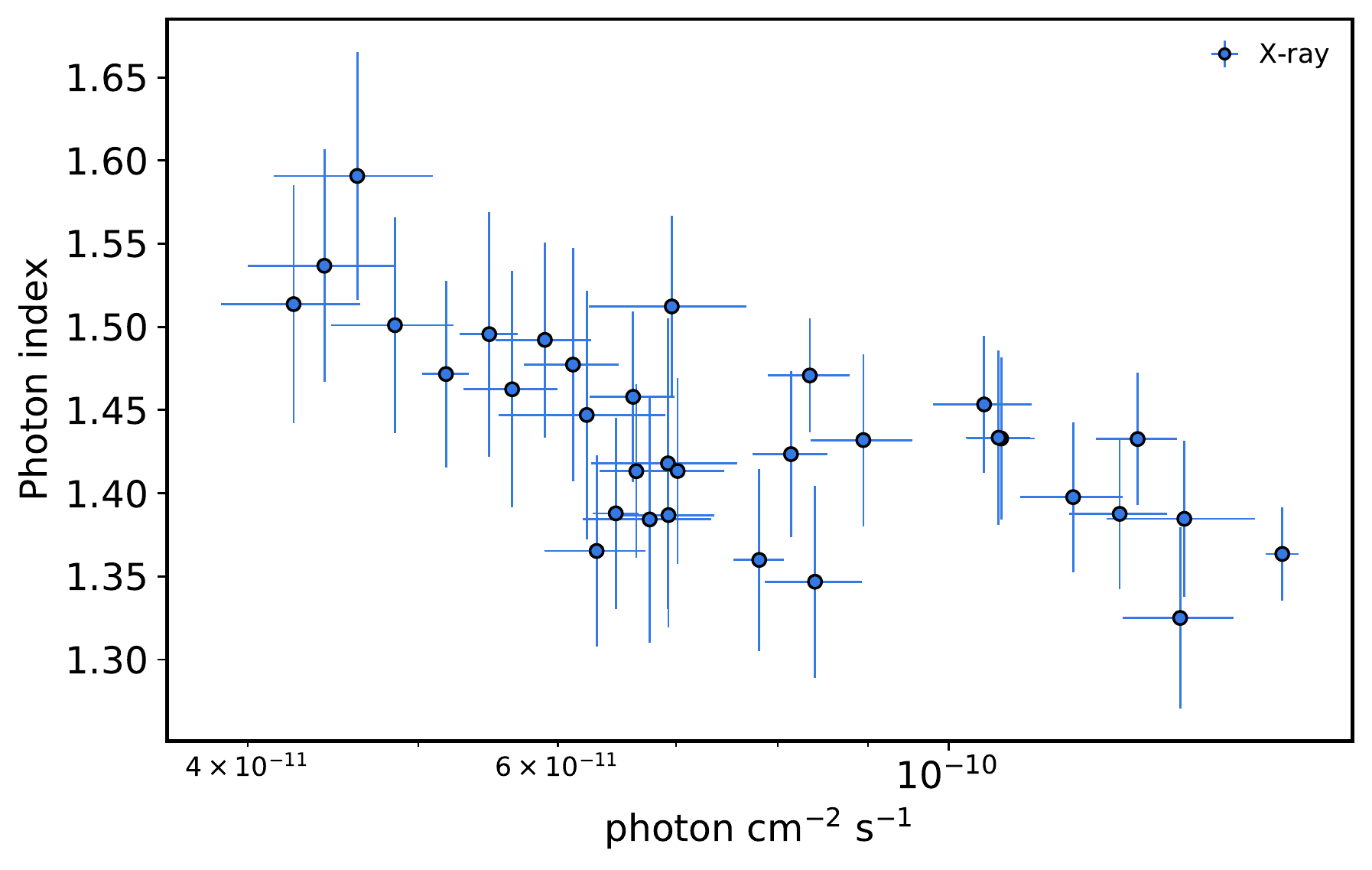}
    \caption{Photon index versus flux. {\it Left panel:} The \gray\ photon index vs. the flux during the major \gray\ flare (MJD 55400-55800) estimated in 3-day bins. {\it Right panel:} The X-ray photon index vs. the flux during MJD 55400-55600.}
    \label{index}
\end{figure*}
\indent The evolution of the \gray\ photon index in the considered ten years was also investigated. Unlike the fast \gray\ flux variation, the photon index does not vary significantly. It is hard to investigate the photon index variation when adaptive bins are considered, as the time intervals are short and the photon index is estimated with large uncertainty; the simple $\chi^2$ test results in $\chi^2/dof\simeq1.01$ with a probability of $P(\chi^2)=0.12$, implying a relatively constant photon index. Instead, in the light curve with 3-day bins, the accumulation time is enough for precise estimation of the photon index showing that it varies as well.  Considering only the intervals when TS $>$ 16, the hardest photon index of $1.87\pm0.04$ was observed on MJD $56559.16\pm1.5$ with $47.5\sigma$. The source flux was $(2.54\pm 0.15)\times10^{-6}\:{\rm photon\:cm^{-2}\:s^{-1}}$, implying the source was in an average \gray\ emitting state. Furthermore, the \gray\ photon index correlation was investigated.
The correlation of the photon index and the flux is expected when the accelerated electrons are cooled down; the photon index is controlled by acceleration or cooling times, and depending on which of them is dominating, the photon index at the peak flux either softens or hardens \citep{1998A&A...333..452K}. Photon index hardening is observed for some blazars and radio galaxies when they get brighter \citep[e.g., see][]{2010ApJ...710..810A, 2010ApJ...721.1425A, 2010ApJ...721.1383A, 2017ApJ...848..111B}. In the long time scale, any trend will be smoothed out because of the mix of different flux levels, so the spectral variability and flux correlation was investigated using the data around the major \gray\ flare (MJD 55400-55800). The \gray\ photon index versus flux estimated in 3-day bins is presented in Fig. \ref{index} (left panel) and the possible correlation is investigated using the linear-Pearson correlation test. The test yielded $r_{p}=-0.41$, implying a negative correlation between the flux and photon index, i.e., the flux increases with decreasing (hardening) photon index. 
It should be noted that a very moderate harder when brighter trend was measured using daily bins for the period covering the outbursts in 2009 December and 2010 April \citep{2010ApJ...721.1383A}.
\subsection{Swift XRT}
The Swift satellite with three instruments on board, the UV and Optical Telescope \citep[UVOT, ][]{2005SSRv..120...95R}, the X-Ray Telescope \citep[XRT, ][]{2005SSRv..120..165B} sensitive to the 0.3-10.0 keV band, and the Burst Alert Telescope \citep[BAT,][]{2005SSRv..120..143B}
sensitive to the 15-150 keV band is an ideal instrument for simultaneous observation of blazars in the X-ray, Optical, and UV bands. \source\ was monitored by Swift 465 times during 2005 - 2016. In the current study the data collected by XRT and UVOT instruments has been analyzed.\\
\indent The XRT data were taken both in photon counting mode (PC) or windowed timing mode (WT) with the single exposure ranging from 0.22 to 14.35 ks for a total exposure of $\sim0.76$ Ms. All the XRT data were processed using {\it Swift$\_$xrtproc} which is an automatic tool for XRT data analysis developed within the Open Universe initiative \footnote{https://openuniverse.asi.it} (\citet{GiommiOU}, Giommi et al. 2021, submitted). The raw event files  (Level1) were reduced, calibrated and cleaned via the XRTPIPELINE script by applying the standard filtering criteria and the latest calibration files of CALDB. The source counts were extracted from a circular region of a radius of $\sim20$ pixels ($47^{''}$) centered on the position of \source, while the background counts are taken from an annular ring centered at the source, with inner and outer radii of $\sim51$ pixels ($120^{''}$) and $\sim85$ pixels ($200^{''}$), respectively. For some observations, the source count rate was above $0.5\:{\rm counts\:s^{-1}}$ and the data were significantly affected by the pile-up in the inner part of the point-spread function. These pile-up effects were removed by excluding events within the circle the radius of which is defined by the count rate and varying within $3$-$6$-pixels. The ungrouped data were loaded in XSPEC (vesiopn 12.11) for spectral fitting using Cash statistics \citep{1979ApJ...228..939C}. The individual spectra were fitted adopting absorbed PL and log-parabola models with the galactic absorption column density of $6.78\times10^{20}\:{\rm cm^{-2}}$ \footnote{https://heasarc.gsfc.nasa.gov/cgi-bin/Tools/w3nh/w3nh.pl}.\\
\indent Fig. \ref{lightcurve_all} d) shows the 2.0-10.0 keV X-ray flux variation during 2007-2016. The X-ray variation over different observations is evident, the lowest flux being $(7.47\pm0.59)\times10^{-12}{\rm erg\:cm^{-2}\: s^{-1}}$ and the highest $(1.80\pm0.18)\times10^{-10}{\rm erg\:cm^{-2}\: s^{-1}}$. The source was observed many times around the major \gray\ brightening, showing the X-ray flux increased between MJD 55517-55520 when in seven observations the flux was above $\sim10^{-10}{\rm erg\:cm^{-2}\: s^{-1}}$. 
Another period of bright X-ray emission was observed in MJD 55166-55174 when again the X-ray flux was $>10^{-10}{\rm erg\:cm^{-2}\: s^{-1}}$. The X-ray flux varies albeit with lower amplitude also during the \gray\ flaring period in MJD 56800-57800.\\
\indent Fig. \ref{lightcurve_all} e) shows the variation of X-ray photon index measured in the $0.3-10$ keV band. The X-ray photon index is around $\sim1.5$ and does not show substantial variation over different observations. However, occasionally harder or softer indexes are observed; the softest index of $1.75\pm 0.04$ was observed on MJD 56829.60 while the hardest, $1.16\pm 0.20$, on MJD 54831.70. 
The photon index variation versus the flux is investigated using the data around the flares at MJD 55130-55250 and MJD 55400-55600. In the latter case, the linear-Pearson correlation test shows no correlation, while a negative correlation ($r_{p}=-0.60$) is found during the first flare (Fig. \ref{index} right panel). Along with the increase of the flux, the photon index hardens to $\sim1.35$.
\begin{table*}
\caption{Results of the \gray\ spectrum fitting with different models (PLEC and LP).}\label{params}
\centering
\begin{tabular}{llccccc}
\hline\hline
Period & Flux$^1$ & $\Gamma/\alpha$ & $E_{\rm cut}^2$ & $\beta$ & $\sigma$ & $\sqrt{2(\Delta\mathcal{L})}$\\
\hline
54719.82 -  54730.50 & $  3.94 \pm0.44$ & $ 2.15 \pm0.05$ & $4.03 \pm 1.00$ & - - &  93.31 &   31.47 \\
54808.37 -  54850.18 & $  0.82 \pm0.05$ & $ 2.21 \pm0.06$ &    $3.61 \pm 0.96$ & - - &  63.71 &   26.53 \\
54955.18 -  54990.41 & $  0.85 \pm0.12$ & $ 2.13 \pm0.06$ &    $3.20 \pm 0.81$ & - - &  60.30 &   30.53 \\
55051.41-$55065.48^*$   & $1.27\pm0.05$ &	 $2.14\pm0.07$ & - -  &	$0.18\pm0.04$ &	52.42 &	26.09\\
55104.72 -  55118.41 & $  3.10 \pm0.09$ & $ 2.25 \pm0.06$ &    $3.13 \pm 0.79$ & - - &   79.08 &   30.29 \\
55160.50 -  55166.10 & $  7.84 \pm0.20$ & $ 2.13 \pm0.04$ &    $5.05 \pm 1.18$ & - - &   95.78 &   35.89 \\
55177.89-$55182.82^*$ &	$6.94\pm	0.22$ &	$2.08\pm	0.04$& - - &	$0.11\pm	0.02$& 94.54&	31.36 \\
55213.93 -  55223.36 & $  5.06 \pm0.13$ & $ 2.22 \pm0.05$ &    $3.34 \pm 0.76$ & - - &   85.59 &   37.13 \\
55246.42 -  55264.45 & $  3.06 \pm0.07$ & $ 2.25 \pm0.05$ &    $3.83 \pm 0.94$ & - - &   96.03 &   29.57 \\
55280.79 -$55288.24^*$ &   $5.79 \pm0.19$ &   $2.19 \pm0.04$ & - - &   $0.11 \pm0.02$ &   92.52 &         25.52 \\
55303.44 -  55308.62 & $  7.48 \pm0.19$ & $ 2.17 \pm0.05$ &    $4.11 \pm 1.08$ & - - &   88.65 &   26.90 \\
55371.82 -  55414.87 & $  1.53 \pm0.04$ & $ 2.20 \pm0.05$ &    $2.75 \pm 0.52$ & - - &   88.15 &   52.14 \\
55414.87 -  55422.58 & $  2.71 \pm0.30$ & $ 2.07 \pm0.09$ &    $1.67 \pm 0.43$ & - - &   59.08 &   28.70 \\
55464.76 -  55480.30 & $  2.29 \pm0.38$ & $ 2.03 \pm0.08$ &    $1.70 \pm 0.35$ & - - &   68.19 &   34.55 \\
55480.30 -  55494.20 & $  3.71 \pm0.41$ & $ 2.27 \pm0.05$ &    $4.06 \pm 1.03$ & - - &   98.09 &   30.11 \\
55494.20 -$55500.87^*$ &   $5.80 \pm0.18$ &   $2.23 \pm0.05$ & - - &    $0.12 \pm0.02$ &   87.34 &         28.15 \\
55502.06 -  55510.51 & $ 10.54 \pm0.17$ & $ 2.13 \pm0.02$ &   $10.55 \pm 2.13$ & - - &  143.97 &   45.78 \\
55517.05 -  55518.06 & $ 48.96 \pm1.03$ & $ 2.01 \pm0.03$ &    $8.81 \pm 1.97$ & - - &  143.71 &   40.02 \\
55518.06 -  55518.68 & $ 62.37 \pm1.27$ & $ 2.11 \pm0.03$ &    $6.28 \pm 1.42$ & - - &  146.21 &   37.83 \\
55519.59 -  55520.19 & $ 73.59 \pm1.79$ & $ 2.09 \pm0.04$ &    $6.40 \pm 1.68$ & - - &  128.32 &   26.71 \\
55520.19 -  55520.80 & $ 52.29 \pm1.34$ & $ 1.98 \pm0.04$ &    $6.91 \pm 1.57$ & - - &   88.00 &   39.00 \\
55521.30 -$55523.00^*$ &  $24.94 \pm0.72$ &   $2.10 \pm0.04$ & - - &    $0.09 \pm0.02$ &  114.86 & 27.45 \\
55523.00 -  55525.35 & $ 21.23 \pm0.45$ & $ 2.14 \pm0.04$ &    $6.76 \pm 1.65$ & - - &  104.34 &   32.77 \\
55529.04 -  55530.56 & $ 20.31 \pm0.57$ & $ 2.13 \pm0.04$ &    $5.82 \pm 1.61$ & - - &   74.78 &   25.08 \\
55541.35 -  55542.68 & $ 14.88 \pm0.48$ & $ 2.09 \pm0.06$ &    $3.75 \pm 1.03$ & - - &   85.26 &   25.91 \\
55543.94 -  55545.30 & $ 17.02 \pm0.49$ & $ 2.08 \pm0.05$ &    $4.23 \pm 1.10$ & - - &   94.92 &   28.50 \\
55549.00 -  55553.46 & $ 18.28 \pm0.30$ & $ 2.15 \pm0.02$ &    $9.17 \pm 1.83$ & - - &  131.45 &   47.50 \\
55587.20 -  55628.06 & $  2.93 \pm0.07$ & $ 2.30 \pm0.03$ &    $5.76 \pm 1.15$ & - - &  119.14 &   44.66 \\
56657.96 -  56744.44 & $  0.95 \pm0.03$ & $ 2.24 \pm0.08$ &    $4.69 \pm 2.05$ & - - &   75.02 &   28.73 \\
56829.85 -  56832.09 & $ 11.95 \pm0.31$ & $ 1.84 \pm0.03$ &   $14.94 \pm 3.54$ & - - &  114.77 &   37.68 \\
56857.26 -$56863.64^*$ &   $2.12 \pm0.15$ &   $1.89 \pm0.08$ & - - &  $0.22 \pm0.05$ &   40.56 &         27.77 \\
57267.66 -  57278.13 & $  2.84 \pm0.16$ & $ 2.12 \pm0.06$ &    $3.06 \pm 0.80$ & - - &   72.78 &   27.99 \\
57287.48 -$57301.85^*$ &   $2.98 \pm0.10$ &   $2.23 \pm0.05$ & --& $0.12 \pm0.03$ &   81.17 & 25.05 \\
57326.85 -$57333.30^*$ &   $3.25 \pm0.16$ &   $2.06 \pm0.07$  & - - & $0.19 \pm0.04$ &   60.43 &         28.34 \\
57400.81 -  57409.39 & $  4.33 \pm0.12$ & $ 1.99 \pm0.05$ &    $2.90 \pm 0.54$ & - - &   88.21 &   56.42 \\
57431.33 -$57456.50^*$ &   $1.26 \pm0.21$ &   $2.78 \pm0.10$ & - - &  $0.16 \pm0.03$ &   60.63 &         29.99 \\
57558.20 -  57562.21 & $  7.40 \pm0.38$ & $ 1.83 \pm0.07$ &    $4.67 \pm 1.34$ & - - &   55.20 &   26.14 \\
57562.21 -  57562.71 & $ 16.42 \pm0.76$ & $ 1.77 \pm0.08$ &    $3.86 \pm 1.11$ & - - &   64.92 &   27.05 \\
57563.01 -  57564.33 & $ 17.18 \pm0.50$ & $ 1.87 \pm0.04$ &    $5.52 \pm 1.17$ & - - &   89.99 &   47.61 \\
57607.08 -  57635.05 & $  2.37 \pm0.21$ & $ 2.20 \pm0.04$ &    $5.41 \pm 1.24$ & - - &   97.19 &   35.48 \\
57743.77 -  57763.90 & $  1.64 \pm0.05$ & $ 2.13 \pm0.05$ &    $3.66 \pm 0.88$ & - - &   76.37 &   34.49 \\
\hline
$^1$ The flux in units of $10^{-6}\:{\rm photon\:cm^{-2}\:s^{-1}}$.\\
$^1$ The cut-off energy in GeV.
\end{tabular}
\end{table*}
\subsection{Swift UVOT}
Simultaneously with XRT, \source\ was observed with the UVOT instrument in six filters, V (500-600 nm), B (380-500 nm), U (300-400 nm), W1 (220-400 nm), M2 (200-280 nm) and W2 (180–260 nm). All the single observations were processed. The source counts were extracted from a circular region of $5^{''}$ around the source, while the background - from a $20^{''}$ region away from the source not containing any significant pixel. The source magnitudes and fluxes were extracted using {\it uvotsource}. Magnitudes are corrected for the galactic extinction \citep{ 2011ApJ...737..103S} 
which were then converted to fluxes using the central wavelength values for each filter from \citet{2008MNRAS.383..627P}.\\
\indent Fig. \ref{lightcurve_all} panels f) and g) show the light curves in optical and UV bands (separating V, B, U and W1, M2 and W2 filters for clarity). On the low state the optical/UV flux of the source is at the level of $(7-8)\times10^{-12}\:{\rm erg\:cm^{-2}\:s^{-1}}$ which increases above $\sim3\times10^{-11}\:{\rm erg\:cm^{-2}\:s^{-1}}$ during the flares. The flux increased in all six filters with different amplitudes: the highest flux of $(1.16\pm0.03)\times 10^{-10}\:{\rm erg\:cm^{-2}\:s^{-1}}$ was observed on MJD 53507.85 before the \fermi\ operation (not shown in Fig. \ref{lightcurve_all}). This coincides with the exceptional optical outburst observed in spring 2005 \citep{2006A&A...453..817V}. The flux increased almost to the same level, $(1.03\pm0.03)\times 10^{-10}\:{\rm erg\:cm^{-2}\:s^{-1}}$, also in $B$ band on MJD 56832.89. During the \gray\ major brightening around MJD 55500 and the flare in MJD 56800-57800, the UV flux in all bands substantially increased, being above $\sim5\times10^{-11}\:{\rm erg\:cm^{-2}\:s^{-1}}$.\\
\indent The visual inspection of the light curves shown in Fig. \ref{lightcurve_all} suggests that the flares in different bands are nearly correlated or appear with small legs. A cross correlation analysis of the \source\ emission in different bands has been performed in \citet{2021ApJ...906....5A} where a detailed comparison of the emission in various bands is provided \citep[see Table 2 and 3 in][]{2021ApJ...906....5A}.
\begin{figure*}
	\includegraphics[width=0.47\textwidth]{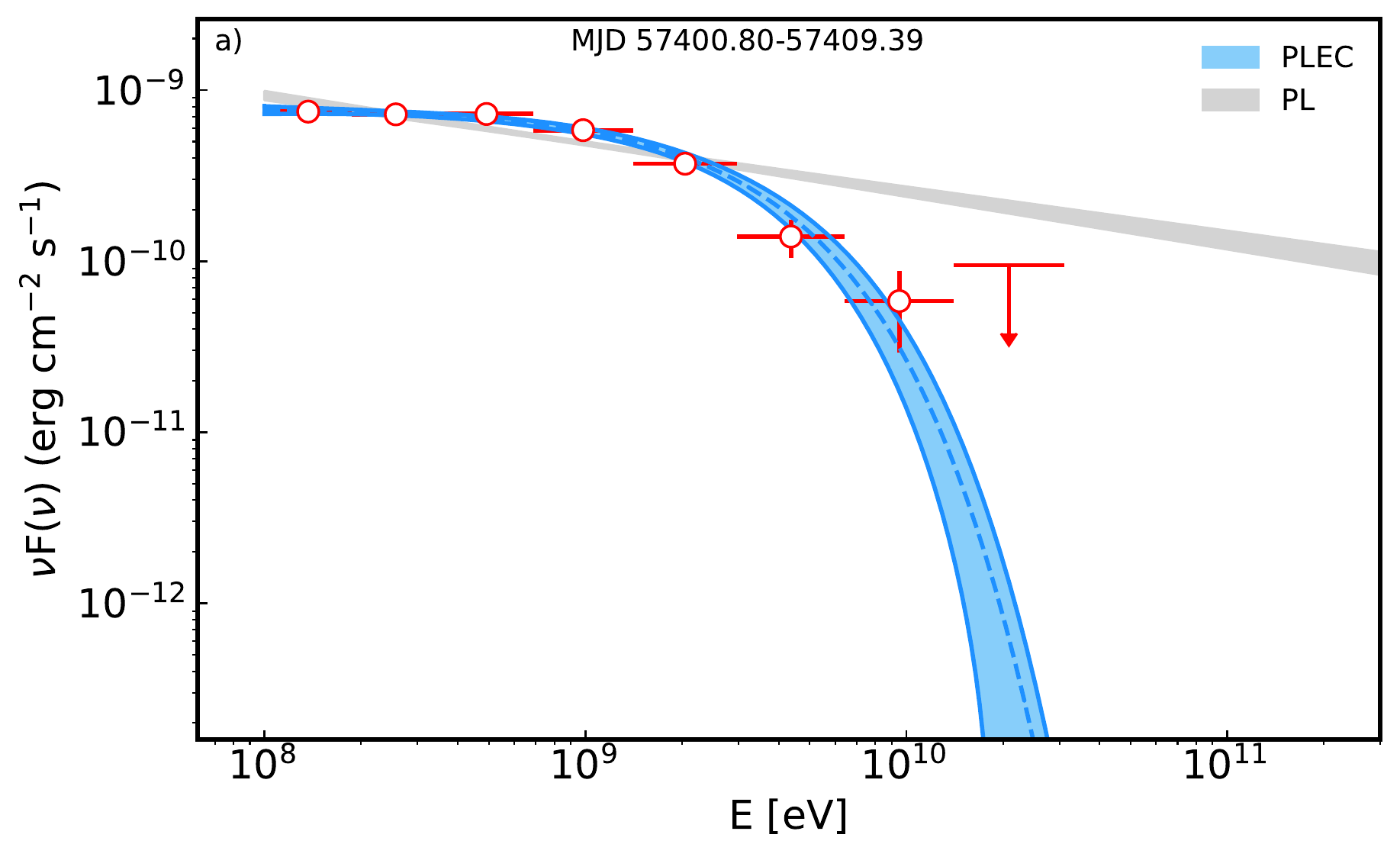}
	\includegraphics[width=0.47\textwidth]{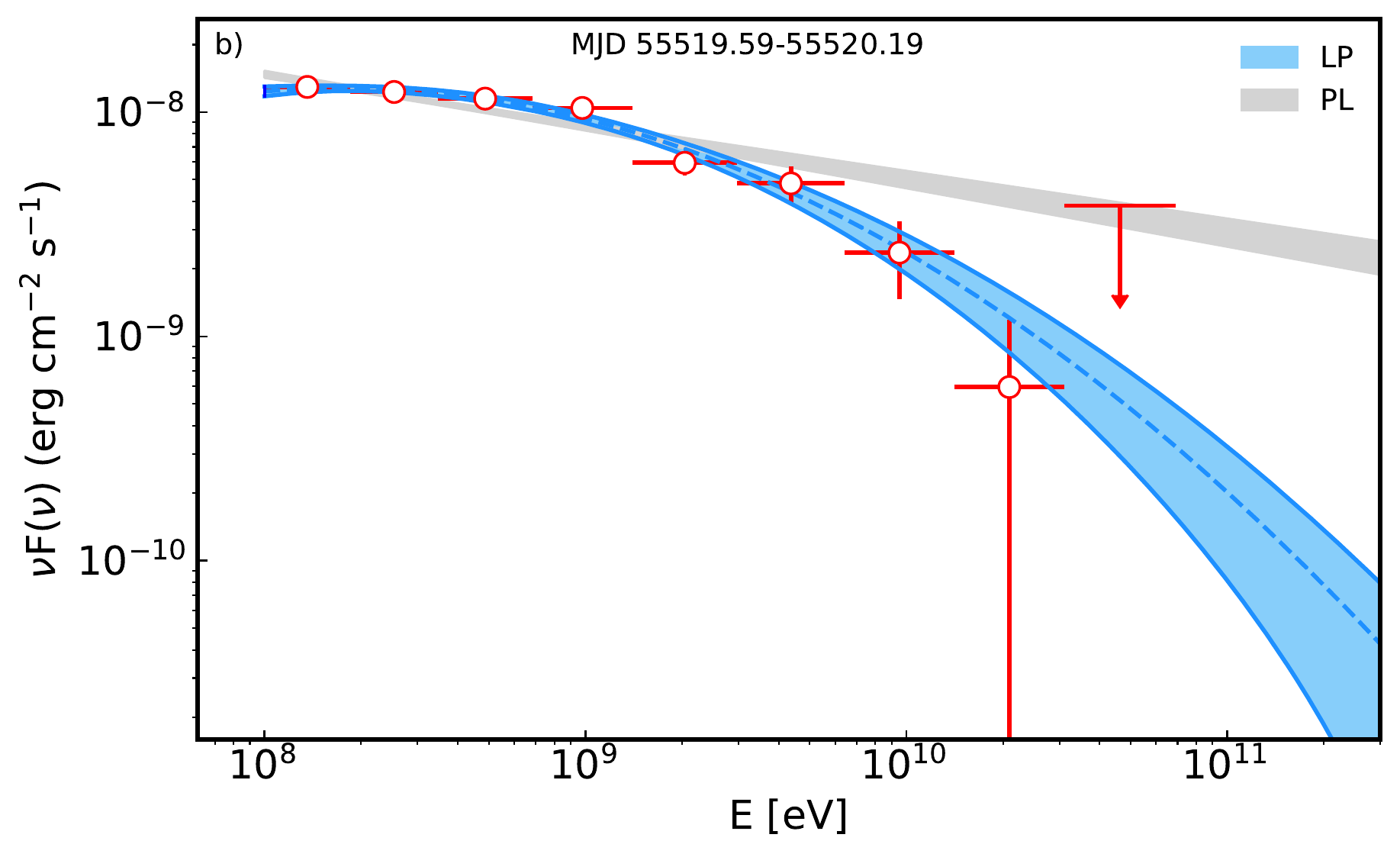}\\
	\includegraphics[width=0.47\textwidth]{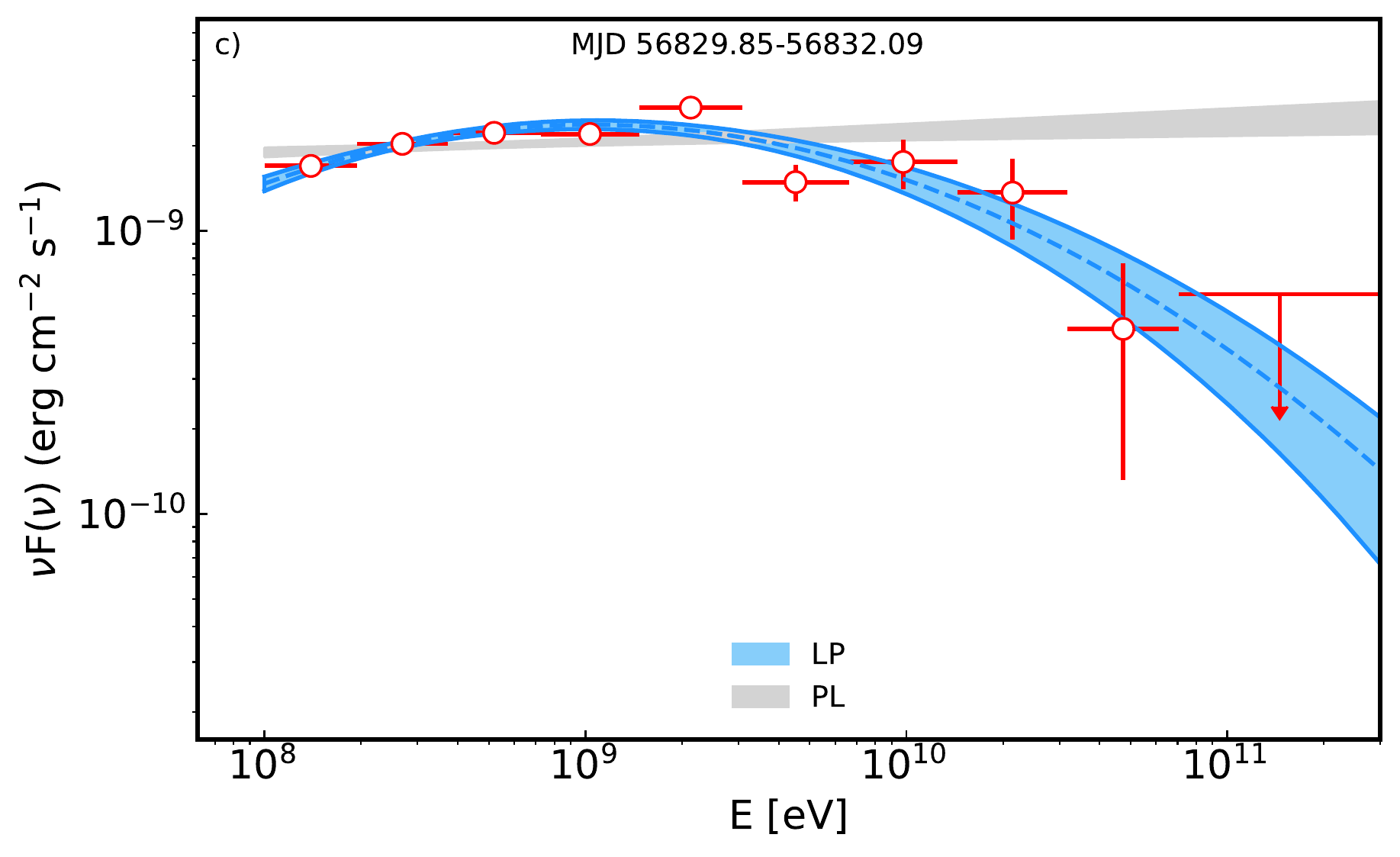}
	\includegraphics[width=0.47\textwidth]{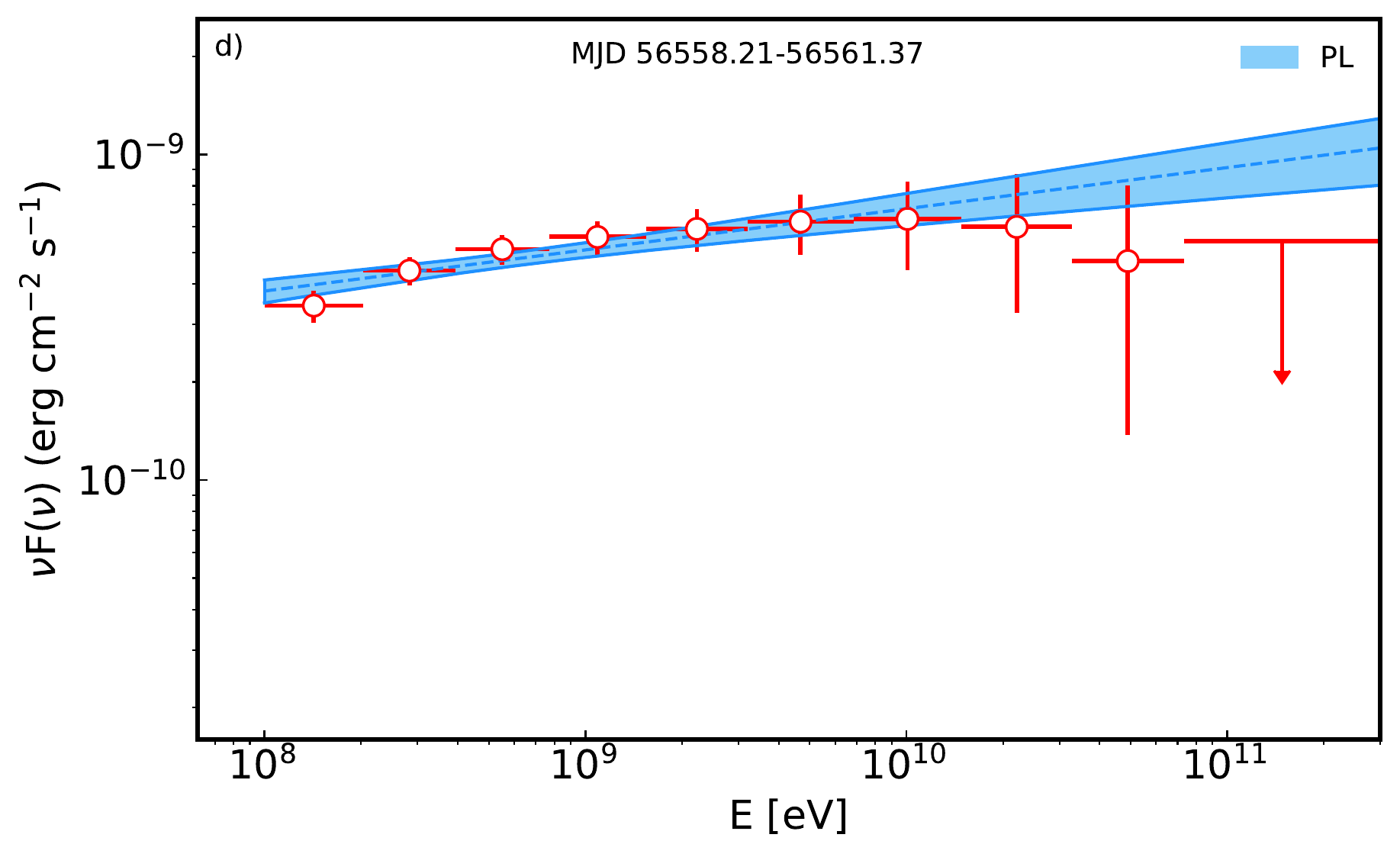}\\
    \caption{The \gray\ SEDs of \source\ in different periods. Panels a) and b) PLEC and LP spectral modeling (blue shaded area) versus PL model (gray shaded area), respectively. Panels c) and d) the periods with hard \gray\ spectra. The spectral points have been obtained by running {\it gtlike} tool for smaller energy intervals.}
    \label{curvature}
\end{figure*}

\section{Evolution of Spectral Energy Distribution}\label{sedevol}
The data presented in the previous section provide a detailed view of the long-term emission of \source. The temporal evolution of the SEDs is investigated by generating SEDs with simultaneous or quasi simultaneous data. The SEDs are constructed in the following manner: for each interval the \gray\ data are plotted together with the Swift UVOT, XRT or, if available, archival data extracted from the ASI Space Science Data Centre (ASI/SSDC) \footnote{https://tools.ssdc.asi.it/SED/}. The archival data observed both at low and high frequencies are included, allowing to constrain the SEDs from $10^{6}$ Hz to $10^{26}$ Hz.
The intervals are selected based on the \gray\ data as the source is being continuously monitored since 2008. Yet, in the ideal case, the \gray\ spectral points should be generated for all adaptively binned intervals and compared with the Swift observations. However, for short intervals, in the \gray\ band the spectrum will extend only up to moderate energies of $\sim1$ GeV not enough for theoretical modeling. So, in order to overcome the problem of low statistics, the adaptively binned light curve is divided into piecewise constant blocks (Bayesian blocks) by optimizing a fitness function \citep{2013ApJ...764..167S}. This gives the optimal segmentation of the data into time intervals during which the data are statistically consistent with a constant flux. These blocks provide an objective way to detect significant local variations in the light curve. In this way, the \gray\ emitting intervals with the same flux level (weather flaring or constant) will be selected and separated. These intervals are shown in Fig. \ref{lightcurve_all} c panel. By this statistical method, the selected intervals will be longer, allowing to calculate the \gray\ spectra up to reasonable energies necessary for theoretical modeling.\\
\indent The Bayesian block algorithm applied to the adaptively binned light curve produces 388 intervals each with a constant flux level. Similarly, Bayesian blocks are computed also for the 3-day binned light curve, which in general produces similar results although with less intervals. In order to have a more detailed view of the SEDs evolution, the blocks from adaptively binned light curve are considered. The spectrum in each Bayesian block is computed by applying unbinned likelihood analysis assuming the spectrum of \source\ is a PL with the normalization and index as free parameters. Then, the SEDs are calculated by fixing the source PL index and running {\it gtlike} separately for 4 to 7 energy bins (depending on the source significance) of equal width in log scale.\\
\indent The resultant SED evolution in time (SED/ligh curve animation) can be found in \href{https://www.youtube.com/watch?v=wNLVj3W6ZFg}{\nolinkurl{youtube.com/wNLVj3W6ZFg}} showing dramatic changes in the broadband spectrum of \source\ during 2008-2018. The flux amplification is nearly of two orders of magnitude in the optical/UV bands while in the X-ray band it is of two-three orders of magnitude. The major changes are observed in the \gray\ band when the flux changes by four orders of magnitude. Similar changes of the flux in the optical/UV and X-ray bands show that, perhaps, in these bands the same component is contributing, whereas in the \gray\ band another component is dominating. This fits well in the synchrotron and SSC/EIC scenario for the broadband emission from \source.
\begin{figure*}
	\includegraphics[width=0.47\textwidth]{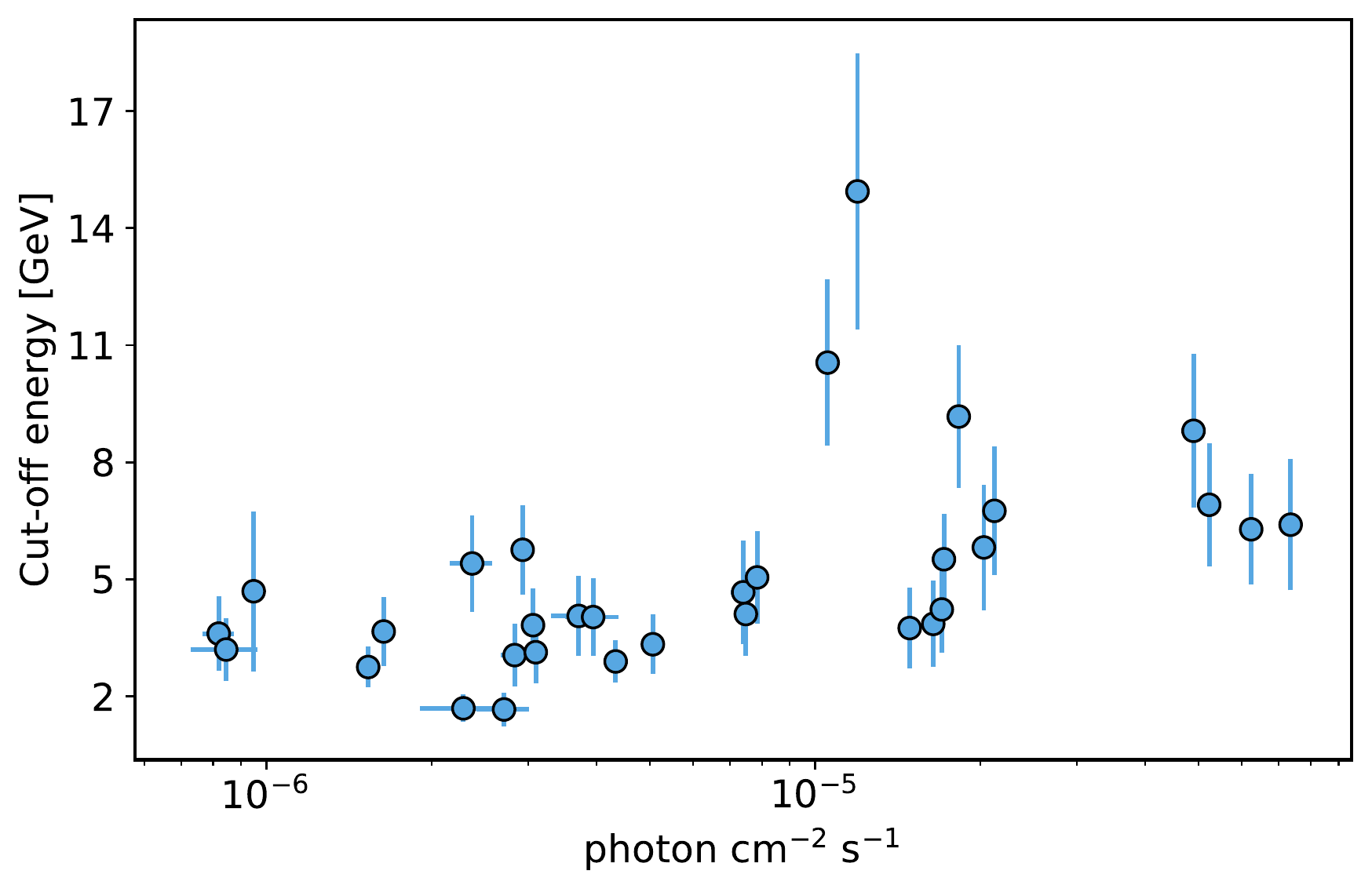}
	\includegraphics[width=0.47\textwidth]{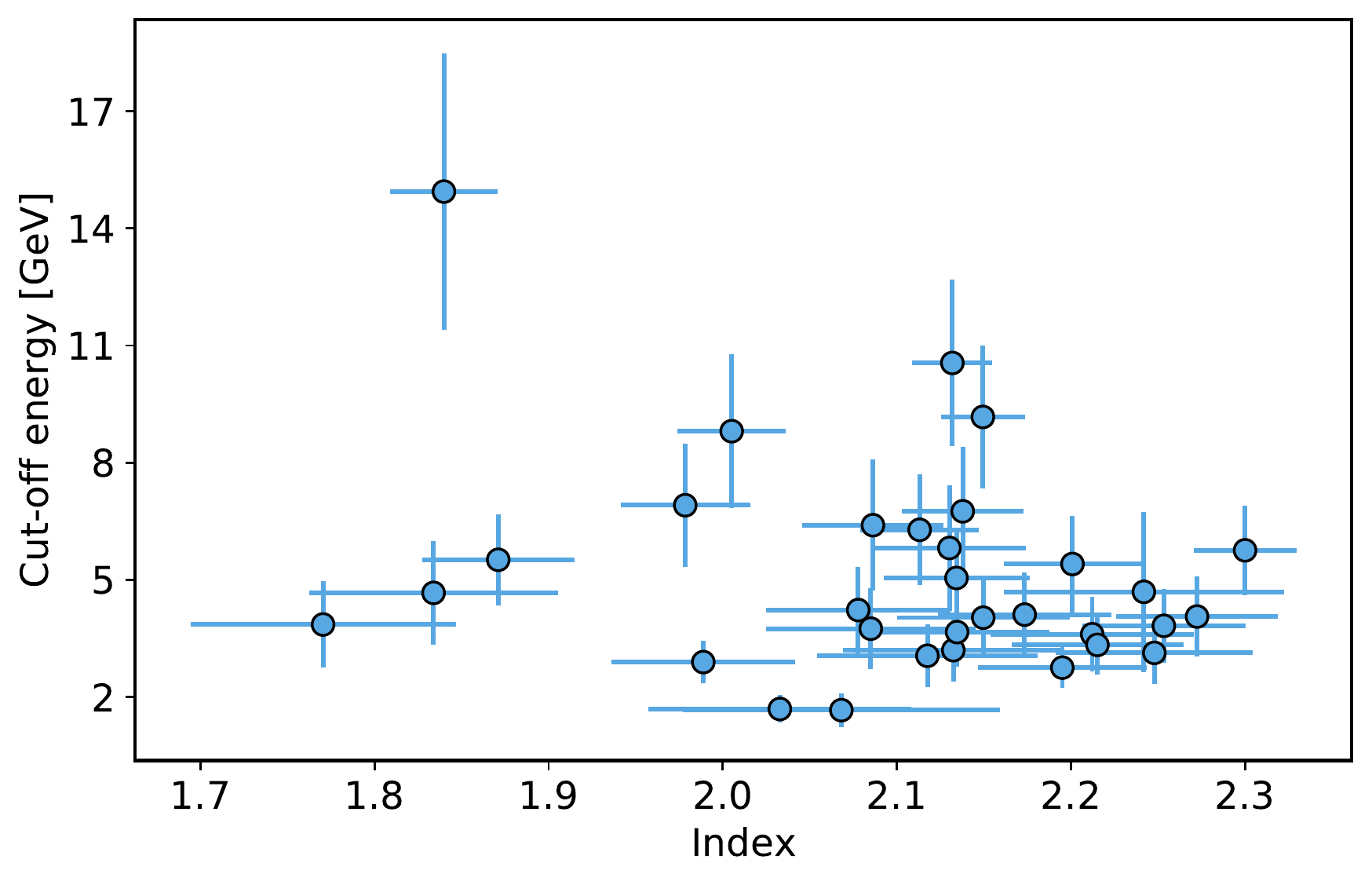}
    \caption{The cut-off energy versus the \gray\ flux (left panel) and photon index (right panel).}
    \label{phcut}
\end{figure*}
\subsection{Gamma-ray spectrum evolution in time}
The periods when the \gray\ spectrum deviates from the simple PL model (red bowtie) can be identified in the time evolution of the multiwavelength SEDs. In order to identify whether the curvature is statistically significant, alternative fits with functions in the form of $dN/dE\sim E_{\gamma}^{-\alpha}\:\times Exp(-E_{\gamma}/E_{cut})$ (power-law with exponential cut-off [PLEC]) and $dN/dE\sim (E_{\gamma}/E_{\rm br})^{-(\alpha+\beta log(E_{\gamma}/E_{\rm br}))}$ (log-parabola [LP]) were applied. These models are compared with PL modeling by applying a log likelihood ratio test where the significance is $2(\mathcal{L_{\rm PLEC/LP}}-\mathcal{L_{\rm PL}})$. In the 388 Bayesian blocks, there are 41 intervals when the significance of the curvature was $>5\sigma$. These periods are given in Table \ref{params} for each interval, providing the \gray\ flux, photon index, cut-off energy or $\beta$ if the model is LP, detection significance and curvature significance.\\
\indent In all the periods, both PLEC and LP provide a statistically better representation of the data and only in nine periods (marked with $*$) only LP modeling was preferred over the simple PL model. In all the periods, the detection significance of the source was $>40\:\sigma$, convincingly high to test the curvature. In these periods the \gray\ spectrum is soft with a photon index of  $1.98-2.30$, except for five intervals when it was $<1.90$.  Fig. \ref{curvature}  a) and b) panels show the \source\ spectra during two periods (with a high significance of the curvature) where the PLEC and LP are compared with PL model. Interestingly, curved spectra were observed also when the source was in active and hyperactive states (MJD 55502-55554 in Table \ref{params}). The spectrum in the hyperactive state with a flux of $(7.36\pm0.18)\times10^{-5}\:{\rm photon\:cm^{-2}\:s^{-1}}$ is shown in Fig. \ref{curvature} b).\\ 
\indent The cut-off energy variation versus the \gray\ flux and photon index is shown in Fig. \ref{phcut}. No strong cut-off energy variation is found as compared to the flux, i.e., the flux varies by a factor of $\sim90$, whereas the cut-off energy only by $\sim 5$. Similarly, as the photon index occasionally can be as hard as $\simeq1.77$, the break energy remains relatively constant. This is in agreement with the previous studies of \source\ 
\citep[e.g.,][]{2011ApJ...733L..26A, 2010ApJ...721.1383A}. The highest cut-off energy of $E_{\rm cut}=14.95\pm3.54$ GeV has been observed on MJD 56829.85-56832.09 when the source was in a bright \gray\ emission state with a flux of $(1.19\pm0.03)\times10^{-5}\:{\rm photon\:cm^{-2}\:s^{-1}}$ and the \gray\ photon index was $1.84\pm0.03$.\\
\indent Interestingly, there are periods when the \gray\ spectrum substantially hardened shifting the peak of the second emission component towards higher energies. The spectrum of \source\ in such two periods in shown in Fig. \ref{curvature} panels c) and d). In the first period (panel c), even if the \gray\ spectrum is initially hard, it starts to curve above a few GeV, and the LP model better explains the data. On the other hand, the \gray\ spectrum measured during MJD 56558.21-56561.37 is with a photon index of $1.84\pm0.04$ and extends above $\sim10$ GeV with $51.50\sigma$. Additional periods with harder \gray\ photon index of $1.88\pm0.06$, $1.89\pm0.07$ and $1.89\pm0.06$ were observed during MJD 56834.20-56834.91, 56916.17-56916.64 and 56827.09-56827.52, respectively. Although, these are short intervals as compared with the period in Fig. \ref{curvature} d) and the spectra were measured up to $\sim10$ GeV, the hardening of the spectrum is evident.
\section{Origin of multiwavelength emission}\label{ome}
\begin{figure}
	\includegraphics[width=0.47\textwidth]{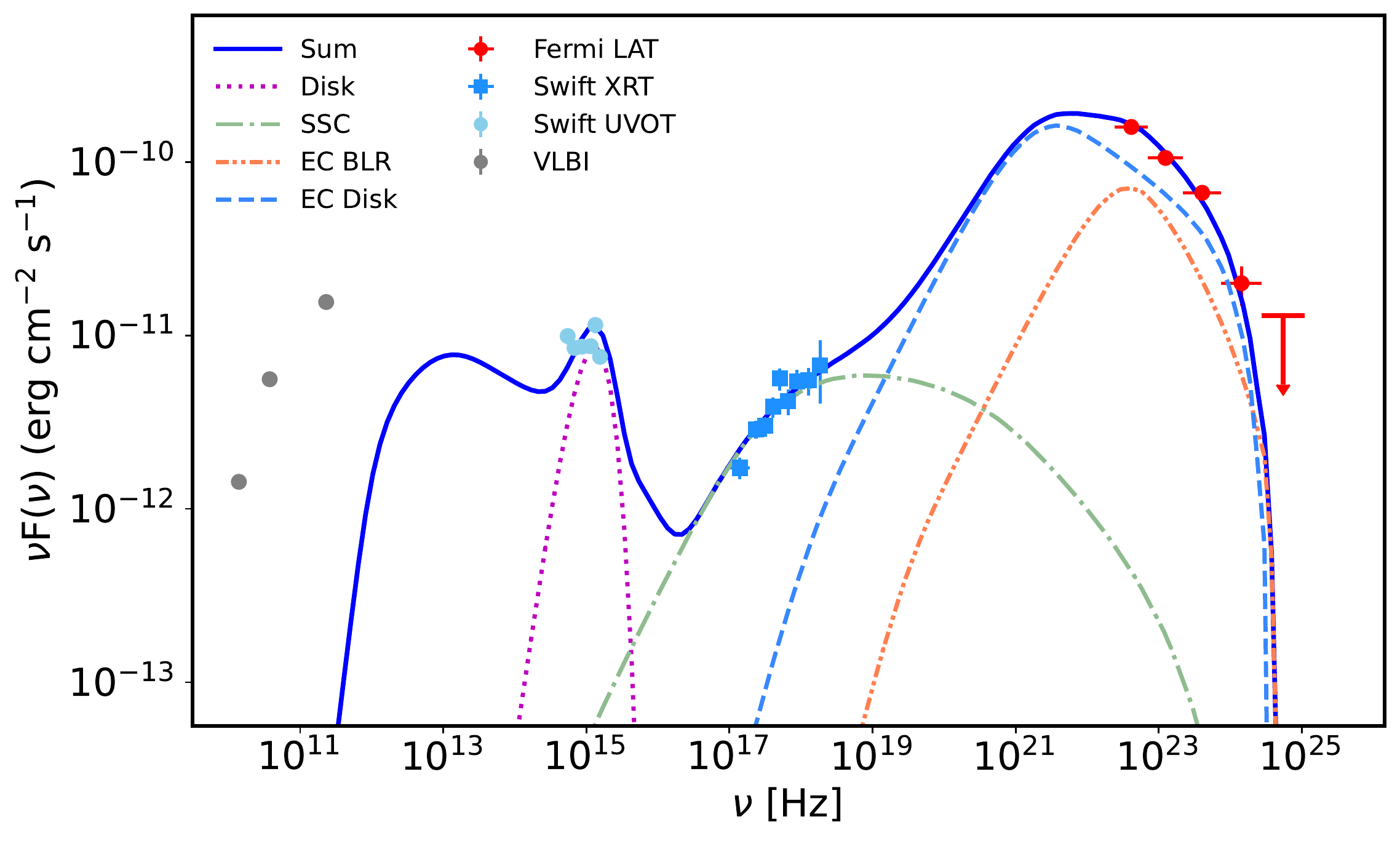}\\
	\includegraphics[width=0.47\textwidth]{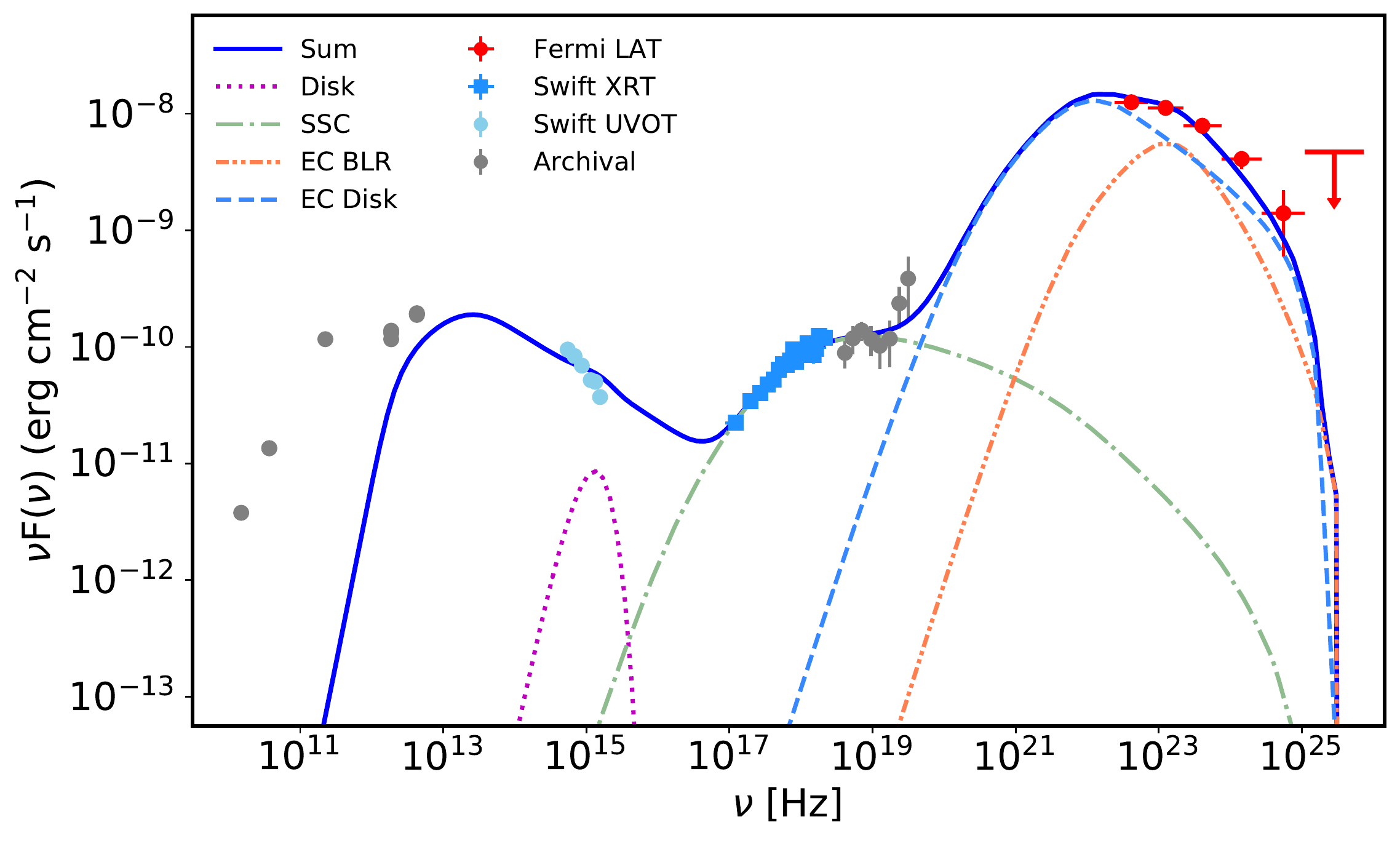}\\
	\includegraphics[width=0.47\textwidth]{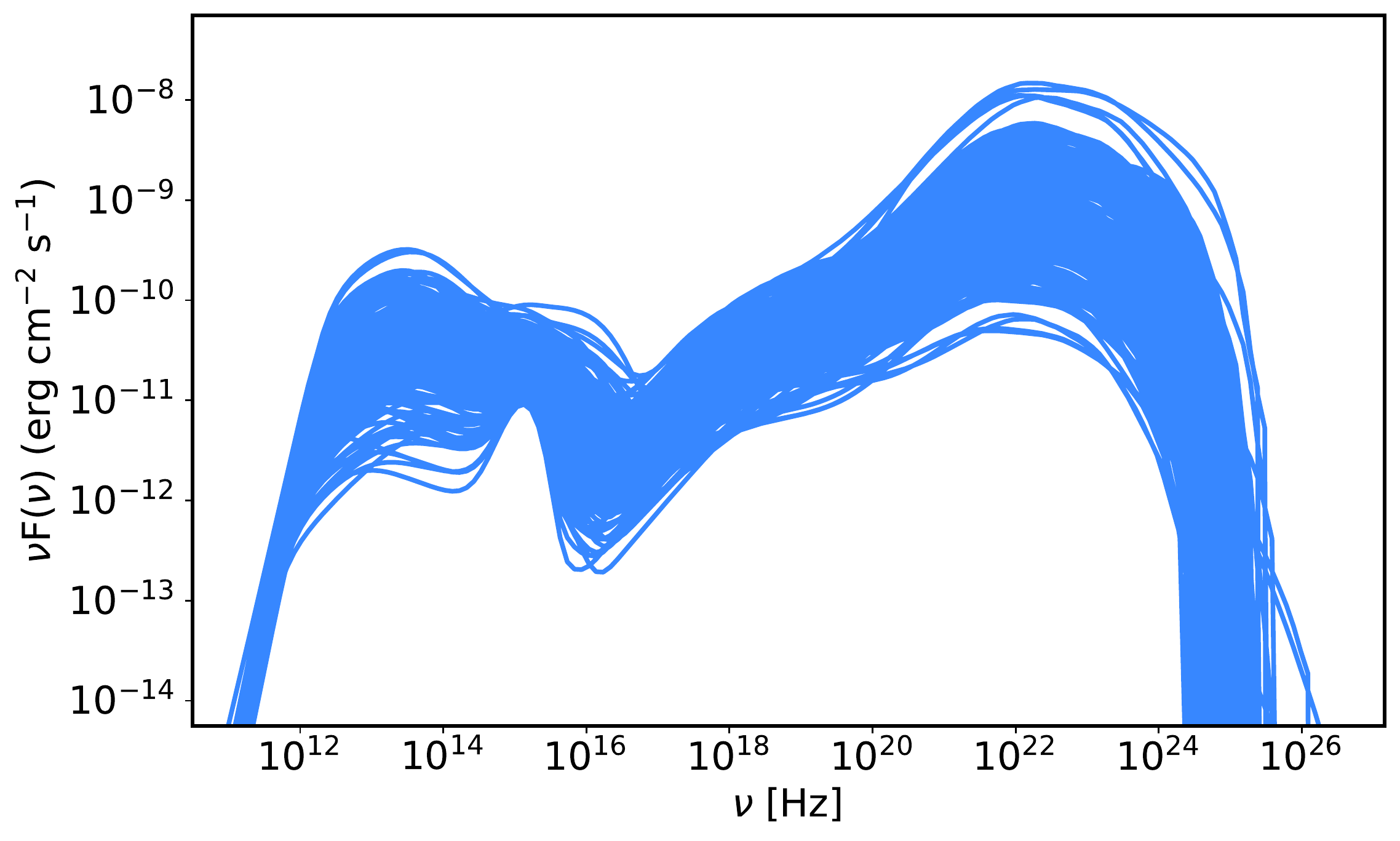}
    \caption{{\it Upper panel:} The multiwavelength SED during MJD 54808.37-54750.83. \fermi, Swift XRT and UVOT data are in red, blue and light blue, respectively. The VLBI radio data are from \citet{2010ApJ...712..405V}. {\it Middle panel:} The multiwavelength SED during the bright period in MJD 55519.59-55520.19. {\it Lower panel:} The summary of all components from the modeling of all SEDs with contemporaneous data collected during 2008-2018.}
    \label{low_sed}
\end{figure}

The multiwalength data obtained in Section \ref{datanal} provide unprecedented detailed information on the emission spectrum of \source\ over different years. Yet, the large amount of the available data allows to investigate not only the emission in different states but also, through theoretical modeling, the evolution of different components of the SEDs. By modeling single snapshot SEDs constrained by (quasi) contemporaneous data, the main parameters describing the jet can be estimated, whereas modeling of the SEDs of the same source observed in different periods can provide a clue on the changes in the jet over different periods. Such an interpretation of the data is the backbone of any model aiming to self-consistently explain blazar emission. Thus, in all the periods shown in the SED/light curve animation when the data in the optical/UV, X-ray and \gray\ bands are available (362 periods) have been modeled and the corresponding parameters estimated.\\
\indent The double-peaked SED of \source\ is modeled within a homogeneous one-zone leptonic scenario where the low energy component is interpreted as synchrotron emission of relativistic electrons, while the second component is due to inverse Compton up-scattering of various photon fields. The seed photons come from the jet itself \citep[SSC model e.g.,][]{maraschi, bloom} and those from the accretion disk \citep[external Compton scattering of direct disk radiation, EC disk;][]{1992A&A...256L..27D} or those reflected from the BLR clouds \citep[EC BLR;][]{sikora} or those of the dusty torus \citep{blazejowski}, depending on the location of the emission region along the jet. However, the observed high energetics of \source\ as well as the short time variability in the \gray\ band suggest that the emission region is located close to the blazar central black hole where the dominant photon fields are those from the accretion disk and BLR. Similar assumption was made in \citet{2010ApJ...712..405V}, and \citet{2010ApJ...714L.303F} demonstrated that the combination of those two photon fields can explain the sharp break in the SED of \source\ and it gives a better fit to the quasi-simultaneous radio, optical/UV, X-ray and \gray\ data.\\
\indent In this scenario, the emission is produced in a spherical blob of the jet with a size of $R$ filled with uniform magnetic field $B$ that moves with a Lorentz factor of $\Gamma$ at a small angle to the observer. The emission from the blob is enhanced by $\delta\simeq\Gamma$ for small angles. The emission region is filled with a population of non-thermal electrons which have a broken power-law distribution in the form of
\begin{equation}
N(\gamma')=\left\{
    \begin{array}{ll}
    N'_e\gamma'^{-p_{1}}    &\mbox{$\gamma'_{\rm {min}}\leq\gamma'\leq\gamma'_{\rm{br}}$}\\
    N'_e{\gamma'_{br}}^{p_{2}-p_{1}}\gamma'^{-p_{2}}   &\mbox{$\gamma'>\gamma'_{\rm{br}}$}\\
    \end{array},
\right.
\end{equation}
where $p_{1}$ and $p_{2}$ are the low and high indexes of electrons correspondingly below and above the break energy $\gamma'_{\rm{br}}$, $\gamma'_{\rm min}$ is the minimum electron energy in the jet frame. $ N'_e$ is connected with the total energy of electrons $U_{\rm e}= m_e\:c^2 \int_{\gamma'_{\rm min}}^{\gamma'_{\rm max}} \!\gamma'\: N(\gamma') \, \mathrm{d}\gamma'$ which scales with the magnetic field energy density $B^2/8\pi$.\\
\indent It is assumed that the emission region is within the BLR, at a distance of $10^{17}$ cm. \citet{2011MNRAS.410..368B} derived that for \source\ the BLR is located at a distance of $\sim6\times10^{17}$ cm. Therefore, the BLR is modeled as a spherical shell with a lower boundary of $R_{\rm in}\simeq4.9\times10^{17}$ cm and an outer boundary of $R_{\rm out}=1.2\times R_{\rm in}\simeq5.9\times10^{17}$ \citep{2003APh....18..377D} and which reflects 10\% of the disk luminosity $L_{\rm disk}$. The disk emission is approximated as a mono-temperature black body.\\
\indent In order to reduce the number of free parameters, it is assumed that the emission region size is $R=5\times10^{15}$ cm which corresponds to hour scale variability as observed in the \gray\ band. Next, in order to constrain the source parameters (e.g., the disk luminosity), the SED observed in MJD 54808.37-54750.83, where the disk contribution in the optical/UV band (blue bump) can be seen, is modeled. The fitting is performed with the open source package {\it JetSet} \citep{2006A&A...448..861M, 2011ApJ...739...66T, 2009A&A...501..879T}.\\
\indent The SED well reproduced by the applied model is shown in Fig. \ref{low_sed} (upper panel). The sum of all components is shown in blue, while the disk component is in magenta peaking at $3.02\times10^4$ K ($6.3\times10^{14}$ Hz) with the luminosity of $L_{\rm disk} = 4.57 \times 10^{46}\:{\rm erg\:s^{-1}}$ similar to the values usually estimated for \source. The X-ray emission is dominated by the contribution from SSC (green) while the EC of disk and BLR components (light blue and red, respectively) dominate at higher energies. The fitting resulted in $\delta\simeq16.9$ which is typical for the bright blazars and $\gamma'_{\rm min}=5.88$ implying that all electrons are cooling in the emitting region. The PL indexes of electrons change from 1.34 to 3.51 at the break energy of $240.7$.\\
\indent When modeling the SEDs in other periods, the disk luminosity and temperature are fixed to the values obtained from the fitting of the SED in MJD 54808.37-54750.83. The other model parameters ($N'_e$, $p_1$, $p_2$, $\gamma'_{\rm min}$, $\gamma'_{\rm br}$, $\delta$ and $B$) have been then estimated by fitting the SEDs. The SEDs modeling animation is available here \href{https://youtu.be/dAqVjpO5Nb4}{\nolinkurl{youtube.com/dAqVjpO5Nb4}}. The synchrotron/SSC components account for the data up to the X-ray band and, in principle, SSC could extend to HEs. However, as demonstrated in \citet{2010ApJ...714L.303F}, the SSC component cannot explain the sharp break observed in the \gray\ band and in such case the strong contribution from BLR photons should be neglected. Considering inverse Compton scattering of only BLR photons, which have a narrower distribution than the SSC component starts to decrease and cannot explain the \fermi\ data. Instead, the HE spectrum can be well reproduced when considering the joint contribution from EC of disk and BLR photons. The EC disk component dominates in the sub-GeV band, while the contribution of EC BLR is significant at HEs. However, the \gray\ data can be explained also when considering the joint inverse Compton scattering of BLR and torus photons \citep[e.g., see ][]{2021arXiv210208962K}.\\
\indent The modeling of the SED during the bright period in MJD 55519.59-55520.19 is shown in Fig. \ref{low_sed} (middle panel). In this active emission state, the source flux from radio to X-ray bands increased nearly by an order of magnitude while in the \gray\ band by nearly two orders of magnitude. As compared with the results of the modeling of the SED in MJD 54808.37-54750.83, the parameters describing the emitting electrons did not vary substantially (e.g., $p_{1}=1.19$, $p_{2}=3.77$ and $\gamma'_{\rm{br}}=271.7$ were estimated in MJD 55519.59-55520.19), but a higher $\delta\simeq50.8$ was estimated. This implies that the flaring activity was caused by the changes in the bulk Lorentz factor of the emitting region.\\ 
\begin{figure*}
	\includegraphics[width=0.99\textwidth]{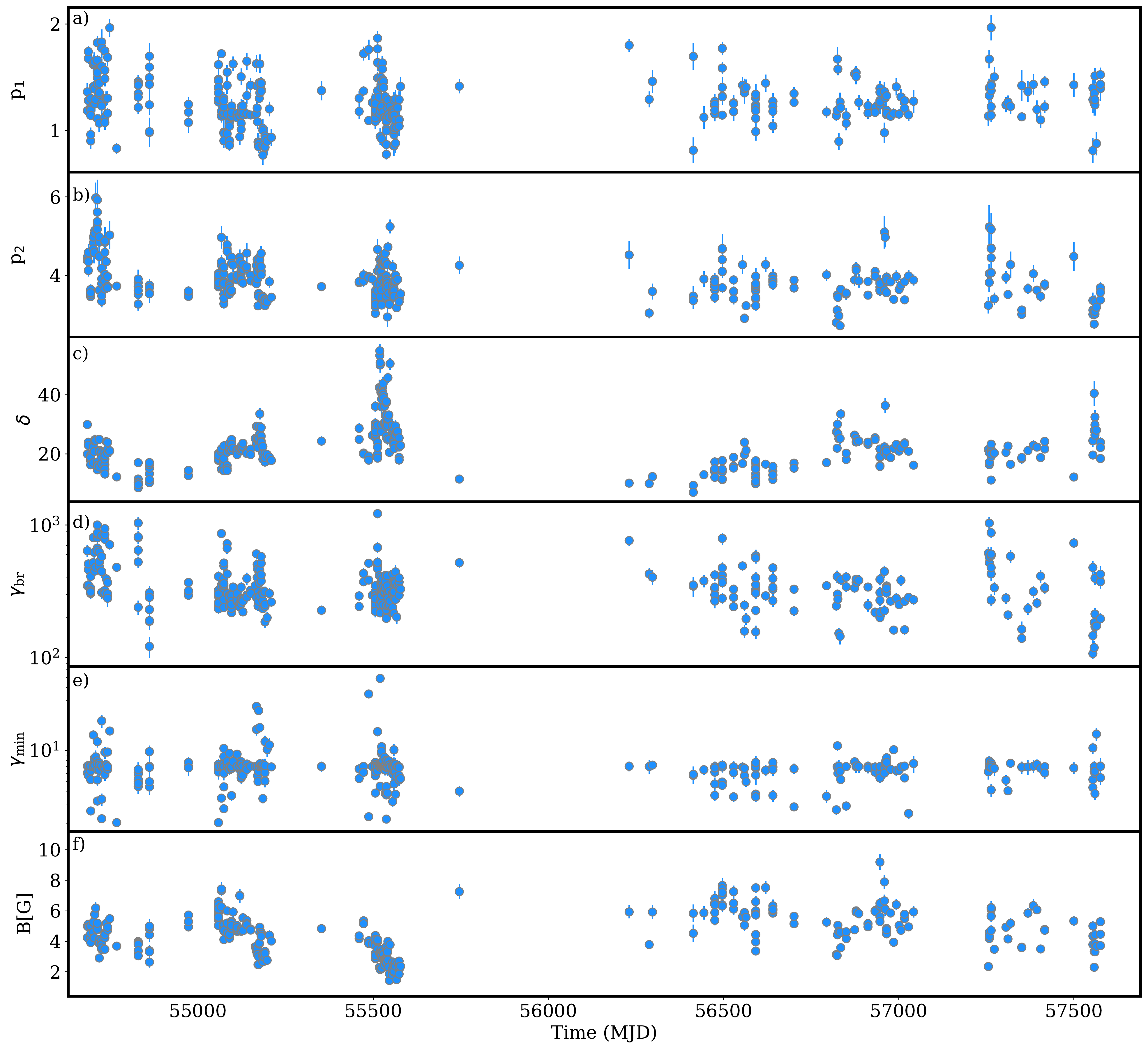}
    \caption{The evolution of model free parameters estimated by modeling the SEDs. {\it a)} and {\it b)} the PL indexes of electrons before and after the break, respectively. {\it c)} the Doppler boosting factor changes in 2008-2018. {\it d)} and {\it e)} the break and minimum energy of emitting electrons in different periods. {\it f)} the change of magnetic field in the emitting region.}
    \label{electparam}
\end{figure*}
\indent The evolution of model parameters is shown in Fig. \ref{electparam}. The photon indexes (panel a and b) are defined by different data sets and vary in the range of $p_{1}=0.77-1.97$ and $p_{2}=2.71-5.98$. The emission in the X-ray band is due to inverse Compton up-scattering of synchrotron photons in the Thomoson regime and the PL index of the emitting electrons is defined by the X-ray data. Similarly, the \gray\ data which are due to inverse Compton scattering of disk and BLR photons are defining $p_{2}$. 
The modeling shows that the $p_1$-$p_2>1.4$, so the index change is significantly larger than that expected from the standard cooling break. The break energy (panel d in Fig. \ref{electparam}) is in the range of $107.07-1220.58$ defining the low and high energy peaks to be at $\sim3\times10^{12}$ Hz and $10^{22}$ Hz, respectively, which is characteristic for FSRQs. The minimum electron energy varies in the range of $2.03$-$48.88$ which is shown in panel e) of Fig. \ref{electparam}.
Along with $p_{1}$ this minimal energy is defined by the SSC modeling of the X-ray data.The magnetic field in the jet is in the range from $1.43$ to $9.19$ G (lower panel of Fig. \ref{electparam}) and the amplitude of its variation is lower than that of the parameters describing the emitting electrons. This implies that the observed flares are likely due to the changes in the emitting electrons rather than in the emitting region plasma. 
\section{Discussion}\label{DC}
The broadband monitoring of \source\ in 2008-2018 reveals an interesting and complicated behaviour in all the considered bands. The highest amplitude flares are observed in the \gray\ band when the flux in several occasions was above $10^{-5}\:{\rm photon\:cm^{-2}\:s^{-1}}$ which corresponds to an apparent isotropic \gray\ luminosity exceeding $10^{50}\:{\rm erg\:s^{-1}}$ (for the 5.49 Gpc distance to \source). In these active states, in the proper frame of the jet, the total power emitted in the \gray\ band would be $L_{\rm em,\gamma}=L_{\rm \gamma}/2\:\delta^2\simeq 10^{47} \: (\delta/20)^{-2}\:{\rm erg\:s^{-1}}$ which by nearly an order of magnitude exceeds the disk luminosity $L_{\rm d}$ in agreement with the results of \citet{2014Natur.515..376G}.\\
\indent The \gray\ photon index of \source\ varies as well; occasionally the photon index can be as hard as $1.87\pm0.04$ measured during a 3-day period. During this period, the highest energy event with $E_{\rm \gamma}=14.92$ GeV has been observed in MJD 56559.89 within a circle of $0.006^\circ$ around \source, with the probability of 0.99987 being associated with it (computed with {\it gtsrcprob} tool). Such hardening of the \gray\ spectrum is unusual for FSRQs which are characterized by a soft \gray\ photon index (e.g., the mean of FSRQ photon index distribution is $2.2$ in 4FGL). However, during the \gray\ flares, occasional hardening of the the \gray\ photon index of FSRQs have been already observed \citep[e.g., see][]{2014ApJ...790...45P, 2018ApJ...863..114G, 2020A&A...635A..25S, 2019ApJ...871..211P, 2019A&A...627A.140A}. 
In some periods the \gray\ spectrum of \source\ deviates from the simple PL model and PLEC and LP models give a better explanation to the overall spectrum. Such modification of the spectrum was observed when the \gray\ emission of the source was in an average or active emitting state, but unlike the strong changes in the flux ($\sim90$ times), the cut-off energy is within $2-10$ GeV.\\
\indent In the X-ray band \source\ behaves like a classical FSRQ with a hard X-ray photon index of $\sim1.5$. Unlike the changes in the X-ray flux, which can increase up to $10^{-10}{\rm erg\:cm^{-2}\: s^{-1}}$, the photon index is relatively constant. The linear-Pearson correlation test reveals a negative correlation ($r_{\rm p}= -0.60$) between the flux and photon index during the flare at MJD 55130-55250. In the optical/UV band \source\ is in an active state after the large outburst in 2005: in several occasions the flux was as high as $10^{-10}\:{\rm erg\:cm^{-2}\:s^{-1}}$. The available optical/UV data allows to shape the peak of the low energy component to be around $10^{13}\:{\rm Hz}$ and unlike the increase of the flux it remains relatively constant. In the past, short transition of the low energy component to higher frequencies during the flares was observed in several FSRQs \citep[e.g., see][]{2002ApJ...571..226C, 2011A&A...529A.145D, 2012MNRAS.421.1764S, 2012MNRAS.420.2899G, 2014MNRAS.445.4316C}. For \source, in this band either the contribution of the disk is observed or, when the synchrotron jet emission dominates during the flares, it corresponds to the falling part of the low energy component which consequently defines the HE tail of the electron distribution. This implies that even during the flares, the processes limiting the maximum energy of the accelerated electrons (e.g., cooling or a limit from the accelerator size) do not change and produce the same effect on the electron acceleration.\\
\indent The multiwavelength SEDs observed in various periods during the considered ten years are well modeled within one-zone leptonic model taking into account the inverse Compton scattering of synchrotron, disk and BLR photons. The adopted model with physically realistic parameters can satisfactorily reproduce the observed SEDs; the models obtained in 362 periods are in Fig. \ref{low_sed} (lower panel) which shows the multiwavelength behaviour of \source\ in 2008-2018. It is evident that the broadband emission varies significantly, except for the radio band which is most likely produced from electrons in more extended regions. This is more evident when comparing the SEDs in the upper and middle panels of Fig. \ref{low_sed} where two different emission states of the source are shown. Unlike the changes in the flux, the peak of both components remains relatively stable. Within the adopted scenario, this could be interpreted by strong cooling of the electrons, i.e., due to the existence of dense photon fields (internal and external), the injected particles cool down, limiting their maximum energy, thus affecting the emission of photons. The different rising and decaying spectra of both components are most likely related with the initial injection (cooling) of electrons.\\
\indent The modeling reveals that the emitting electrons initially are distributed with a hard PL index with a mean of $p_{\rm 1, mean}=1.27$ which substantially hardens to $p_{\rm 2,mean}=3.87$. The break energy varies in the range of $107.08-1220.58$ and is defined by the interplay between the particle acceleration and cooling times. The electron cooling time is defined as 
\begin{equation}
t_{\rm cool}=\frac{3\: m_{e}c \:(1+z)}{4 \sigma_{\rm T}\:u_{\rm tot}^{\prime}\:\gamma_{\rm e}^{\prime}}
\end{equation} 
where $u_{\rm tot}^{\prime}=u_{\rm B}+u_{\rm SSC}+u_{\rm disc}+u_{\rm BLR}$. The densities of disk and BLR photons are constant whereas $u_{\rm B}$ and $u_{\rm SSC}$, which depend on the synchrotron and SSC components, vary in different periods (Fig. \ref{low_sed} lower panel). Thus, the small variation of the break energy (see Fig. \ref{electparam}) is defined by the changes in $u_{\rm B}$ and $u_{\rm SSC}$. Yet, in the vast majority of cases the ratio of $u_{\rm SSC}/u_{\rm B}$ is $\geq1$ implying that the SSC cooling cannot be neglected, so the nonlinear effects are important for the formation of particle distribution, i.e., $\gamma_{\rm br}$ and $p_{2}$. Because of these nonlinear effects the difference between $p_{1}$ and $p_{2}$ is larger than that expected from traditional cooling break ($\Delta p=1$). Alternatively, large $\Delta p$ could be due to the nature of the injection process or due to the inhomogeneities in the source \citep{2009ApJ...703..662R}.\\
\indent In the modeling, some SEDs can be modeled only when considering very hard electron spectra; $p_{1}$ is $0.77-1.97$. This index depends on the combination of the data in the X-ray and \gray\ bands and sometimes because of their hard photon indexes, $p_{1}<1$ is required. In the fast electron cooling regime, a much softer $\gamma^{-2}$ spectrum will be formed below $\gamma_{\rm min}$, whereas in the slow cooling regime the hard spectrum of electrons may be due to their initial injection. However, the PL index of $\leq 1.0$ is challenging for many particle acceleration scenarios. The diffuse shock acceleration of particles can form a spectrum as hard as $-1$ depending on the parameters of the shocks \citep[e.g.,][] {2012ApJ...745...63S}. Similarly, a very hard electron spectrum will be formed in the relativistic magnetic reconnection \citep{2001ApJ...562L..63Z, 2014PhRvL.113o5005G, 2014ApJ...783L..21S, 2016ApJ...816L...8W} only under extremely high magnetization conditions ($\geq\:100$).\\
\begin{figure}
	\includegraphics[width=0.47\textwidth]{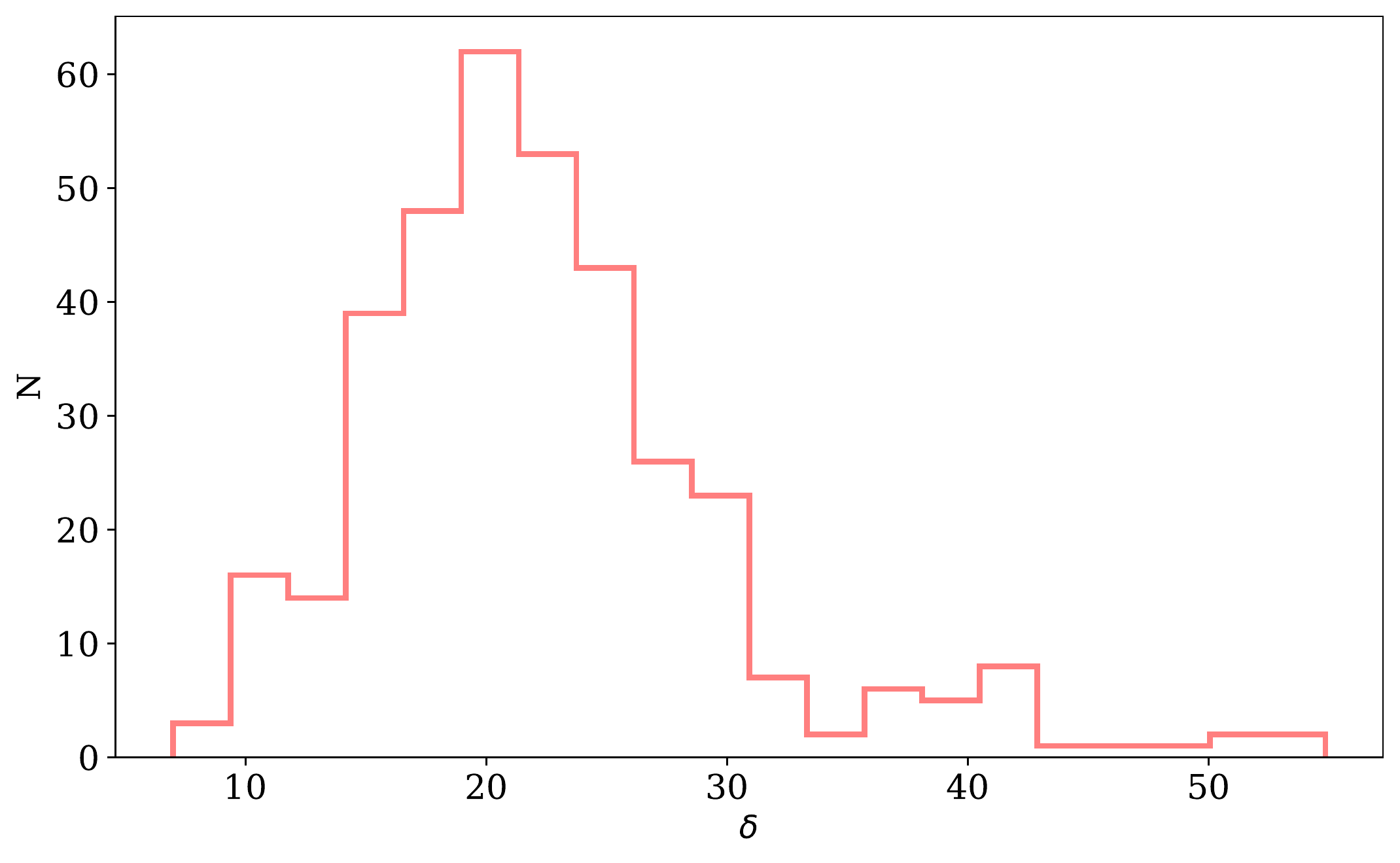}\\
	\includegraphics[width=0.47\textwidth]{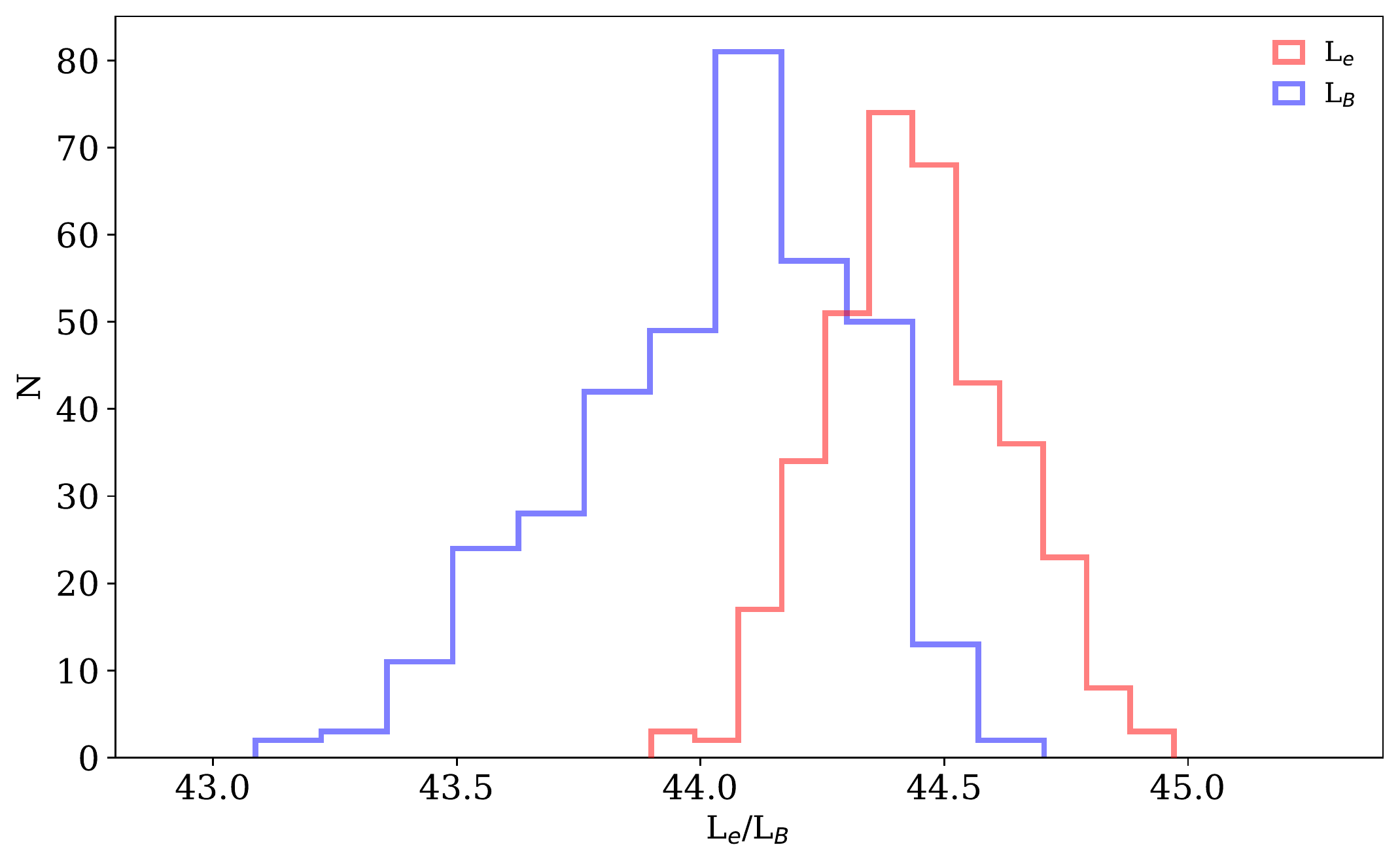}\\
    \caption{{\it Upper panel:} The distribution of Doppler boosting factor. {\it Lower panel:} The distribution of the jet luminosity in the form of electrons (red) and magnetic field (blue).}
    \label{delta}
\end{figure}
\indent The distribution of Doppler boosting factor is presented in the upper panel of Fig. \ref{delta} which has a peak at $\delta\simeq20$ which is characteristic for bright FSRQs \citep{2015MNRAS.448.1060G}. However, in some periods substantially higher values for $\delta$ are estimated: there are 26 periods when $\delta>35$. As can be seen from Fig. \ref{electparam} (panel c), high $\delta$ were estimated around MJD 55500 when the source was in a bright \gray\ emission state. For example, the highest $\delta=54.8$ was estimated in MJD 55518.68-55519.10 when the \gray\ flux was as high as $(5.97\pm0.14)\times10^{-5}\:{\rm photon\: cm^{-2}\:s^{-1}}$ and that in the X-ray band- $(1.59\pm0.07)\times10^{-10}\:{\rm erg\: cm^{-2}\:s^{-1}}$. So, in this period the emission region has a larger Doppler boosting factor which results in a strong increase in the Compton dominance (hence an increase of the HE component) as the external photon density in the comoving frame of the jet depends on the Doppler boosting factor. This faster moving emitting region could be either re-accelerated during the propagation or could be newly injected and emits close to the central source. Another possibility for the Doppler boosting factor increase can be due to geometrical effects, i.e, when the jet regions may have different orientations; e.g., jets in a jet \citep{2009MNRAS.395L..29G} or twisted inhomogeneous jet model \citep{2017Natur.552..374R}. So, during the flares the photons may be produced in a region viewed at smaller angles as compared to the entire jet, which increases the Doppler boosting factor.\\
\indent The modeling provides also information on the power of the jet. The distribution of the jet luminosities in the form of magnetic field and electron kinetic energy computed as $L_{B}=\pi c R_b^2 \Gamma^2 U_{B}$ and $L_{e}=\pi c R_b^2 \Gamma^2 U_{e}$ is given in Fig. \ref{delta} (lower panel). $L_{\rm e}$ is in the range of $(0.79-9.35)\times10^{44}\:{\rm erg\:s^{-1}}$ while $L_{\rm B}$ in $(0.12-5.07)\times10^{44}\:{\rm erg\:s^{-1}}$, implying the system is not far from the equipartition condition. The peak of $L_{\rm e}$ is around $3\times10^{44}\:{\rm erg\:s^{-1}}$ and $L_{\rm B}$ is at $1.2\times10^{44}\:{\rm erg\:s^{-1}}$. For the majority of SEDs, the jet is slightly particle-dominated with $L_{\rm e}/L_{\rm B}\geq1$ and only in a few periods when the low energy components exceed the X-ray flux $L_{\rm e}/L_{\rm B}<1$. The total jet luminosity defined as $L_{\rm tot}=L_{\rm e}+L_{\rm B}+L_{\rm p}+L_{\rm rad}$ \citep{2001MNRAS.327..739G}, where $L_{\rm p}$ and $L_{\rm rad}$ are the power carried by the cold protons and the produced radiation, respectively, is $\leq2.08\times10^{46}\:{\rm erg\:s^{-1}}$, being smaller than the total Eddington luminosity of $1.9\times10^{47}\:{\rm erg\:s^{-1}}$ for the black hole mass of $1.5\times10^9\:M_\odot$ in \source.
\section{Conclusions}\label{conc}
In this paper the broadband emission from \source\ during 2008-2018 is investigated. In this period the source was in active emission state displaying extraordinary flares in the \gray\ band. In several occasions the \gray\ flux exceeded $5\times10^{-5}\:{\rm photon\:cm^{-2}\:s^{-1}}$ corresponding to an apparent isotropic \gray\ luminosity of $>\:10^{50}\:{\rm erg\:s^{-1}}$. Similarly, the source was active (although with lower amplitude) also in the X-ray and optical/UV bands.\\
\indent The multiwavelength SEDs of \source\ (in 362 periods) constrained with contemporaneous data collected during 2008-2018 have been modeled within a one-zone leptonic scenario taking into account the inverse Compton scattering of synchrotron, disk and BLR reflected photons. Through the modeling, the main parameters describing the jet in different periods have been estimated providing an insight into the jet evolution in 2008-2018. It is shown that during the large \gray\ flares the Doppler boosting factor substantially increased which points that the emission during the flares comes most likely from a region which either moves faster or has a different geometrical orientation.
\section*{Acknowledgements}
I thank the anonymous referee for constructive comments. I acknowledge the use of data, analysis tools and services from the Open Universe platform, the Astrophysics Science Archive Research Center (HEASARC) and the Fermi Science Tools.\\
This work was supported by the Science Committee of RA, in the frames of the research project No 20TTCG-1C015.\\
This work used resources from the ASNET cloud and the EGI infrastructure with the dedicated support of CESGA (Spain).
\section*{Data availability}
The data underlying this article will be shared on reasonable request to the corresponding author.



\bibliographystyle{mnras}
\bibliography{biblio} 








\bsp	
\label{lastpage}
\end{document}